\documentclass[final,3p,times,twocolumn]{elsarticle} 

\usepackage{hyperref}
\usepackage{gensymb}  
\usepackage{graphicx}

\usepackage{amssymb}

\usepackage{lipsum}

\usepackage{rotating} 
\usepackage{color} 
\usepackage{subcaption}
\usepackage{graphicx}
\biboptions{comma,square}

\journal{Astroparticle Physics}

\def\mj{{\sc Majorana}}             
\def\dem{{\sc Demonstrator}}             

\begin{document}

\begin{frontmatter}



\title{Solar Axion Search Technique with Correlated Signals from Multiple Detectors}


\author{Wenqin~Xu$^{1,2}$\corref{cor1}}\ead{Wenqin.Xu@usd.edu}
\author{Steven~R.~Elliott$^{1}$\corref{cor2}}

\cortext[cor1]{Corresponding Author}

\address{$^{1}$Physics Division, Los Alamos National Laboratory, Los Alamos, NM, USA}
\address{$^{2}$Department of Physics, University of South Dakota, Vermillion, SD, USA} 

\date{\today}

\begin{abstract}
The coherent Bragg scattering of photons converted from solar axions inside crystals would boost the signal for axion-photon coupling enhancing experimental sensitivity for these hypothetical particles. Knowledge of the scattering angle of solar axions with respect to the crystal lattice is required to make theoretical predications of signal strength. Hence, both the lattice axis angle within a crystal and the absolute angle between the crystal and the Sun must be known. In this paper, we examine how the experimental sensitivity changes with respect to various experimental parameters. We also demonstrate that, in a multiple-crystal setup, knowledge of the relative axis orientation between multiple crystals can improve the experimental sensitivity, or equivalently, relax the precision on the absolute solar angle measurement. However, if absolute angles of all crystal axes are measured, we find that a precision of $2\degree-4\degree$ will suffice for an energy resolution of $\sigma_E=0.04{\rm E}$ and a flat background. Finally, we also show that, given a minimum number of detectors, a signal model averaged over angles can substitute for precise crystal angular measurements, with some loss of sensitivity. 
\end{abstract}

\begin{keyword}
Solar Axions, Solid State Detectors, Dark Matter, Coherent Bragg-Primakoff Conversion
\end{keyword}

\end{frontmatter}




\section{Introduction}
\subsection{Coherent Bragg-Primakoff Conversion of Solar Axions}\label{sec:Primakoff}
The QCD vacuum adds an extra term to the Lagrangian. The magnitude of the term is usually parameterized with an unknown constant denoted as $\theta$. This extra term would violate parity and time reversal symmetries, but would conserve charge conjugation symmetry. As a result, it predicts a non-zero neutron electron dipole moment. Because no such dipole moment has been observed, $\theta$ must be unnaturally small. This situation is referred to as the strong CP problem.

A solution to the strong CP problem is to introduce a spontaneously broken global chiral symmetry. The associated Nambu-
Goldstone boson is referred to as the axion~\cite{Peccei1977,Peccei1977a,Weinberg1978,Wilczek1978a,Peccei2008}. Axions could be the dark matter and if they exist, the Sun would produce copious numbers of them. The science of axions is reviewed in Ref.~\cite{Kim2010}.

The plasma of the solar interior can produce axions through axion-two-photon and axion-electron interactions~\cite{Raffelt1986,Pospelov2008,Gondolo2009,Derbin2011,Redondo2013}. The former is referred to as Primakoff axion production and its strength is given by the coupling constant, $g_{a\gamma\gamma}$. The latter results in atomic-recombination, atomic-deexcitation, Bremsstralhung, and Compton processes that produce axions with rates determined by the axion-electron coupling constant, $g_{ae}$. The flux produced by the Primakoff reaction dominates in hadronic axion models such as the KSVZ~\cite{Kim1979,Shifman1980} where the electron-axion coupling is absent at tree level. The production of axions by the Primakoff process has been studied in Refs.~\cite{Dicus1978,Raffelt1986,Raffelt1988}. The focus of this paper is the detection of axions through coherent Bragg-Primakoff conversion, and although the conclusions developed are applicable to all crystal experiments, the analysis is focused to aid upcoming experiments that will use numerous high purity Ge (HPGe) detectors. The work presented here could also be interesting to other new physics particles that can be produced in the Sun and converted back to photons in detectors, such as the hypothetical solar chameleons~\cite{Brax2010}. 

When solar axions reach a crystal, they can interact with the Coulomb field of nuclei and convert into photons via Primakoff conversion. This process was proposed by Buchm\"{u}ller and Hoogeven~\cite{Buchmuller1990} and Pashcos and Zioutas~\cite{Paschos1994} and developed formally by Creswick {\em et al.}~\cite{Creswick1998}. Given the distance between the Sun and the Earth, solar axions entering each crystal form a parallel beam. If the Bragg condition is satisfied, individual photons from axion Primakoff conversions can coherently sum to produce a strong signal, a process sometimes referred to as coherent Bragg-Primakoff conversion. The rate for this process~\cite{Creswick1998,Cebrian1999} is expressed as, 
\begin{eqnarray}
\frac{dR}{dE_{\gamma}}(t, E_{\gamma}) & = & \int\int \frac{d\sigma}{d\Omega} \frac{d\Phi}{dE_a} \delta(E_a-E_{\gamma})	\\ \nonumber
& & \times~\delta(\vec{k_{\gamma}}-\vec{k_a}-\vec{G})dE_ad\Omega~,
\label{eq:signalrate}
\end{eqnarray}
where $E_a$ is the axion energy and $E_{\gamma}$ is the energy of the axion converted photon.  If the axion mass is much smaller than the axion energy, the outgoing photon carries the same energy as the incoming axion, resulting in the delta function $\delta(E_a-E_{\gamma})$. $\vec{k_a}$ and $\vec{k_{\gamma}}$ are the axion and photon momenta respectively. $\vec{q}\equiv\vec{k_{\gamma}}-\vec{k_a}$ is the momentum transfer, which depends on both the lattice axis angle and the position of the Sun, resulting in a signal strength dependence on time ($t$). $\vec{G}$ is the reciprocal lattice vector $\vec{G}\equiv(h, k, l)\frac{2\pi}{a_0}$, where $a_0=0.566$ nm is the dimension of the unit cell and (h, k, l) are the Miller indices for the crystal axis. $\delta(\vec{k_{\gamma}}-\vec{k_a}-\vec{G})$ arises from the Bragg condition, $\vec{q}=\vec{G}$.  The differential cross section for Primakoff conversion~\cite{Creswick1998} is written as,
\begin{equation}
\label{eq:xsection}
\frac{d\sigma}{d\Omega}=\frac{g_{a\gamma\gamma}^2}{16\pi^2}F_a^2(2\theta)sin^22\theta~,
\end{equation} 
where $2\theta$ is the scattering angle, and thus $q=2ksin(\theta)$. $F_a(2\theta)$ is the form factor of the screened Coulomb field of the nucleus. The theoretical solar axion flux resulting from Primakoff production~\cite{vanBibber1989,Creswick1998} can be written approximately as,
\begin{equation}
\label{eq:flux}
\frac{d\Phi}{dE_a}=\sqrt\lambda\frac{\Phi_0}{E_0}\frac{(E_a/E_0)^3}{e^{E_a/E_0}-1}~,
\end{equation} 
where $\lambda \equiv (g_{a\gamma\gamma}\times10^8~\rm{GeV})^4$, $E_0 \equiv 1.103$ keV and $\Phi_0 \equiv 5.95 \times 10^{14}$ cm$^{-2}$sec$^{-1}$. A more recent parameterization of the axion flux exists~\cite{Raffelt2008, Arik2015} and it can be different from Eq.~\ref{eq:flux} for up to a few percent with almost identical shape. Following the derivation in Refs.~\cite{Creswick1998,Cebrian1999}, the total signal rate of axion-converted photons in a Ge crystal with a volume of V can be expressed,
\begin{eqnarray}
\label{eq:Rate}
\frac{dR}{dE_{\gamma}} & = & (2\pi)^3 2  c \hbar  \frac{V}{(v_c)^2}  \sum_{G} \frac{d\Phi}{dE_{\gamma}} \frac{|S(\vec{G})|^2}{|\vec{G}|^2}  \\ \nonumber
& & \times ~\frac{Z^2\alpha(c\hbar)^2g_{a\gamma\gamma}^2}{16\pi}\frac{q^2(4k^2-q^2)}{(r_0^{-2}+q^2)^2} \\ \nonumber
& & \times ~\delta(E_{\gamma}-\frac{c \hbar |\vec{G}|^2}{2\hat{\vec k} \cdot \vec {G}})~
\end{eqnarray}
where $\hat{\vec k}\equiv\frac{\vec k}{|\vec k|}$ represents the direction of the solar axion flux. $r_0=53$~pm is the screening length in Ge crystals~\cite{WebElements}, and $v_c$ is the volume of the unit cell. $S(\vec{G})$ is the crystal structure function, defined by Eq. 7 in Ref.~\cite{Creswick1998}. A more explicit form for $|S(\vec{G})|^2$ can be found in Ref.~\cite{Avignone2009a}. Ref.~\cite{Creswick1998} has a list of crystal planes relevant for this process with detailed discussions. The remaining delta function results from the Bragg condition expressed in terms of energy. The signal rate depends on the fourth power of $g_{a\gamma\gamma}$, and therefore it is proportional to $\lambda$. 

Following Eq. 5 in Ref.~\cite{Cebrian1999}, to account for finite energy resolution, the experimentally measured photon signal rate can be derived from a convolution,
\begin{eqnarray}
\label{eq:smearing}
\dot R(t, E_{exp})&\equiv& \frac{dR(t, E_{exp})}{dE_{exp}}\\\nonumber&=&\int \frac{1}{\sigma_E \sqrt{2\pi}} e^{\frac{-(E_{exp}-E_\gamma)^2}{2 \sigma_E^2}} \frac{dR}{dE_{\gamma}} dE_\gamma~,
\end{eqnarray}
where $E_{exp}$ is the measured photon energy and $\sigma_E$ is the $1\sigma$ detector energy resolution. This integral removes the delta function in Eq.~\ref{eq:Rate}.

To facilitate calculations, Eq. 10 in Ref.~\cite{Creswick1998} expressed Eq.~\ref{eq:Rate} in a compact form using dimensionless kinematic variables.\footnote{To avoid confusion, we note that Eq. 10 in Ref.~\cite{Creswick1998} has some typographical errors. The constant $\dot{N_0}$ in that equation (or $\dot{N}$ in the text) is incorrectly given as 0.61/kg d. The denominator in $\frac{4\epsilon^2-g^2}{(g^2+\gamma^2)^2}$ should be squared, as stated here. Finally, the definition of $\gamma$ should be $a_0/(2\pi r_0)$.} The compact form with energy resolution incorporated becomes,
\begin{eqnarray}
\label{eq:compact}
\dot R(t, E_{exp}) & = & \frac{M_D \dot{N_0}}{\sigma_E \sqrt{2\pi}}  \lambda \sum_{g} \big[{|S(\vec{g})|^2} \\ \nonumber
& &\times \frac{(4\epsilon_g^2-g^2)}{(g^2+\gamma^2)^2} \times \frac {\epsilon_g^3} {e^{\beta\epsilon_g}-1}  e^{-\frac{(\epsilon-\epsilon_g)^2}{2\epsilon_{\sigma}^2}}\big]~.
\end{eqnarray}
To obtain this compact form, the momenta and energies were made dimensionless with the conversion factor $C_{dim}\equiv\frac{a_0}{c\hbar2\pi}=0.457$~keV$^{-1}$. Explicitly, $\gamma\equiv C_{dim}/r_0$, $\epsilon\equiv E_{exp}C_{dim}$,  and $\epsilon_{\sigma}\equiv \sigma_{E}C_{dim}$. $\epsilon_g\equiv\frac{|\vec{G}|^2}{2\hat{\vec k} \cdot \vec {G}}C_{dim}$ is the dimensionless energy that satisfies the Bragg condition, at which the solar axion flux is evaluated. In natural units ($c=\hbar=1$), $\epsilon_g$ is written as $\epsilon_g=\frac{|\vec{g}|^2}{2\hat{\vec k} \cdot \vec {g}}$, where $\vec{g}\equiv\vec{G}C_{dim}=\vec{G}\frac{a_0}{2\pi}=(h, k, l)$ is the dimensionless version of $\vec{G}$. The specific constants for Ge are $\beta\equiv(E_0C_{dim})^{-1}=1.983$, and $\dot{N_0}=9.504$/(kg d). $M_D$ is detector mass in kg and energy resolution $\sigma_E$ retains its unit of energy, so that the total event rate units are explicit. 

\begin{figure}[h]
\includegraphics[width=0.5\textwidth]{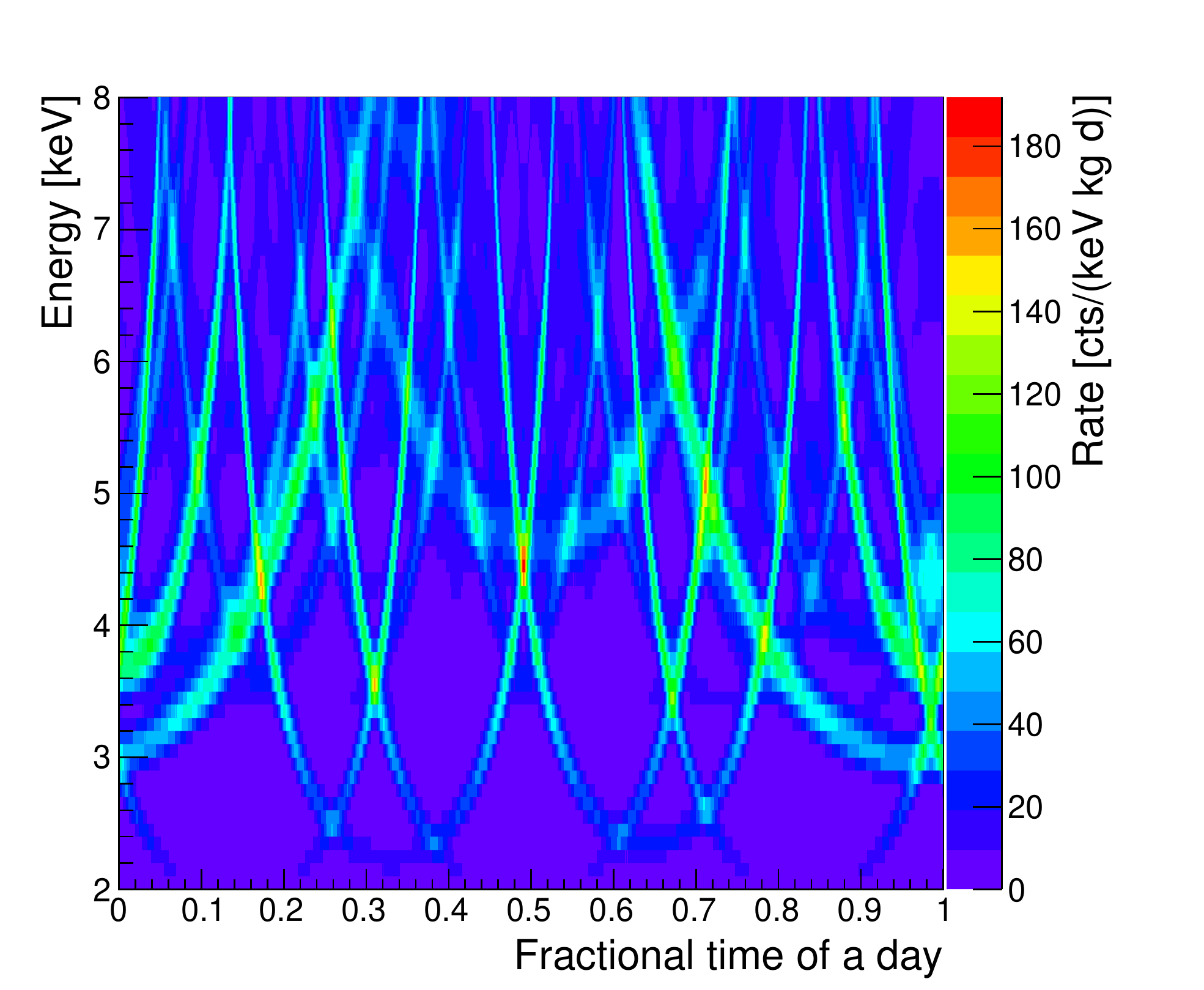}
\caption{Model predicted axion-induced photon signal rate with respect to photon energy and time of a day for $g_{a\gamma\gamma}=10^{-8}$ GeV$^{-1}$ ($\lambda=1$) in a HPGe detector located at Lead, SD with energy resolution $\sigma_E=0.04{\rm E}$.}
\label{fig:PDF}
\end{figure}

\begin{figure}[h]
\includegraphics[width=0.5\textwidth]{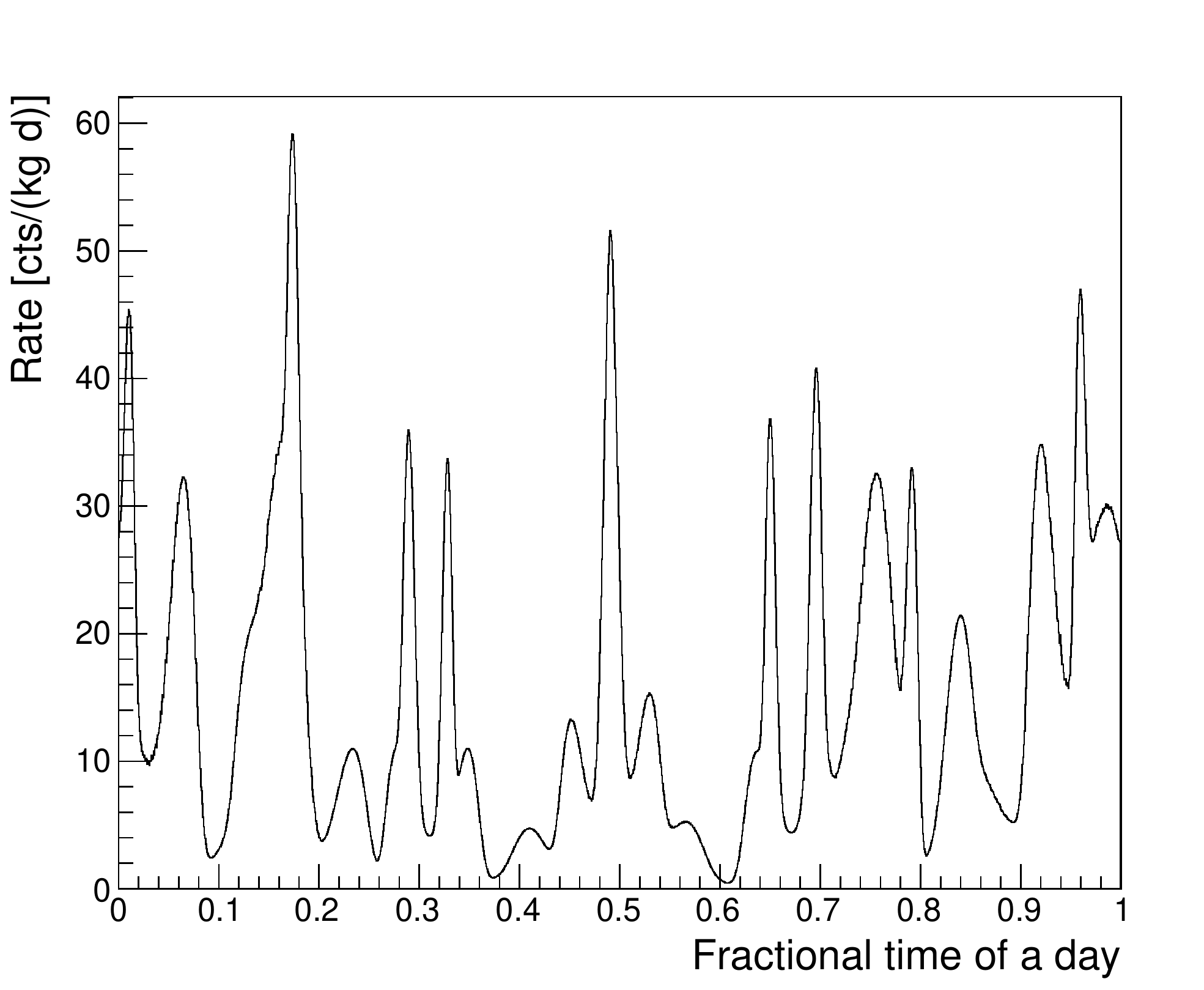}
\caption{Model predicted axion-induced photon signal rate with respect to time of a day with photon energy between 4.0 and 4.5 keV for $g_{a\gamma\gamma}=10^{-8}$ GeV$^{-1}$ ($\lambda=1$) in a HPGe detector located at Lead, SD.}
\label{fig:Energy}
\end{figure}

In an experiment utilizing Bragg scattering, the lattice axis angle is known to some precision and the Sun's location can be obtained based on the time of a day. Eqs.~\ref{eq:Rate} and~\ref{eq:compact} are often used with detected energy and time of day as experimental variables, with axis angle an implicit parameter. Examples of the signal strength as a function of energy and time are shown in Figs.~\ref{fig:PDF} and ~\ref{fig:Energy}. 

\subsection {Experimental Techniques}\label{sec:cystal_axis}
There have been several attempts to detect axions through coherent Bragg-Primakoff conversion. The SOLAX experiment~\cite{Avignone1998} was a pioneer in searching for solar axions using the coherent Bragg-Primakoff conversion. The SOLAX team developed the initial detector-signature phenomenology upon which the Bragg scattering analysis is based. Like SOLAX, the COSME experiment~\cite{Morales2002} used a Ge detector in their search. The CDMS~\cite{Ahmed2009} and EDELWEISS~\cite{Armengaud2013} experiments also used Ge detectors but they were configured as bolometers. The CDMS result is notable due to the well-known orientation of the crystal axis with respect to the Sun. The DAMA experiment~\cite{Bernabei2001} used NaI crystals. Although the exposure of the DAMA measurement is very large compared to the Ge experiments, it has poor energy resolution. The TEXONO~\cite{Chang2007} looked for axions coming from a reactor and interacting coherently. A summary of the solar axion results is given in Table~\ref{tab:ExperSumm}.

 \begin{table}
 \begin{center}
 \caption{A summary of the previous solar axion experimental results using coherent  Bragg-Primakoff conversion.}
 \begin{tabular}{crr}
 \label{tab:ExperSumm}
 Experiment					&	Exposure					& g$_{a\gamma\gamma}$	upper limit	\\
 							&	(kg d)				& $\times 10^{-9}$ GeV$^{-1}$					\\
\hline
SOLAX~\cite{Avignone1998}		&  708\phantom0				& 2.7\phantom0 (95\% CL)		\\
DAMA~\cite{Bernabei2001}		& 53437\phantom0				& 1.7\phantom0 (90\% CL)		\\
COSME~\cite{Morales2002}		& 72.7\phantom0					& 2.78 (95\% CL)		\\
CDMS~\cite{Ahmed2009}			&	443.2\phantom0				& 2.4\phantom0 	 (95\% CL)	\\
EDELWEISS~\cite{Armengaud2013}	& 	448\phantom0				& 2.15 (95\% CL)			\\
\end{tabular}
\end{center}
\end{table}

An assessment of various crystals was studied in Ref.~\cite{Cebrian1999} and the possibility of using large Ge detectors intended for double beta decay was discussed in Refs.~\cite{Avignone2009a,Avignone2009}. Large crystal based experiments being planned for dark matter or double beta decay will have a significant exposure and therefore a substantial improvement in sensitivity for solar axions is expected. For example, Ref.~\cite{Dawei2015} discusses the sensitivity of the CUORE experiment using TeO$_2$ detectors. In Sect.~\ref{sec:Known_Angle} and~\ref{sec:angle_precision} we will examine how the sensitivity of Ge detectors changes with various experimental parameters, including exposure, background, energy resolution and precision on the crystal axis direction relative to the Sun. 

The search for solar axions via coherent Bragg-Primakoff conversion is rarely the primary goal of an experiment. As a result, the crystal axis direction relative to the Sun is frequently poorly known. This was certainly the case for the SOLAX, COSME and DAMA experiments. In the case of CDMS, the detector fabrication process provided additional information that helped define the absolute solar angle to approximately 3$\degree$. To determine the solar angle, the laboratory must be surveyed and the crystal axis must be determined with respect to that survey. Frequently this final step is difficult or is not implemented. However, {\em in situ} source-calibration techniques~\cite{Mihailescu2000,Bruyneel2006,Bruyneel2006a} have the capability of determining the relative angle between different HPGe detectors due to the anisotropy in the charge carrier drift velocity with respect to the crystal axes. As a result, some future experiments may find that the relative angle between detectors is known better than the absolute solar angle. The axion signature in Fig.~\ref{fig:PDF} is complicated but very distinct. Hence, if the relative angles are known, one can exploit that the pattern in one detector will correlate to that in another, as will be shown in Sect.~\ref{sec:Relative_Angle}.

If an experiment utilizes numerous detectors, one can average the signal model and analyze the data without information of crystal azimuthal angles, assuming the crystals are fabricated with [001] axes vertically aligned. The signature of axions within a large number of detectors that are randomly oriented approximates such an average. In Sect.~\ref{sec:averaging}, we consider this approach and estimate the minimum number of detectors that justify this approximation.

\subsection{Simulation Studies}\label{sec:simu_overview}
Axion Primakoff conversion events and background events inside Ge crystals were simulated to form hypothetical experimental runs. Representative signal and background values are chosen to cover the range from the best previous results to that anticipated from near-future experiments. Below, we summarize the values of the parameters chosen for the simulations. Our conclusions are not sensitive to the limited choice of true values within the regions defined by the span of the simulated true values.

The background in the simulations was chosen to be flat in energy and time, and proportional to detector mass. This is a good approximation, because sources of background in this energy range are mainly the smooth Compton continuum of photo-peaks at higher energies. A linear instead of flat background could easily be included. Also, if necessary, a background model incorporating cosmic-ray induced activity and their x ray contributions could be included in the analysis. However, if an experiment takes caution to limit the surface exposure and allows some underground decay time, such as the steps taken in Ref.~\cite{Abgrall2015}, then cosmogenic radioactivity will not significantly distort the nearly flat background. Hence, the added complexity of these background description enhancements were not included and would only slowly change the conclusions in our analysis. The simulations mainly focus on two distinct background levels, $\bar b=$1.5~cts/(keV kg d) and 0.1~cts/(keV kg d), corresponding to a range from what was achieved in recent low background experiments, such as CDMS~\cite{Ahmed2009}, EDELWEISS~\cite{Armengaud2013, Armengaud2016} and the {\sc Majorana Demonstrator} (MJD)~\cite{Elliott2016}, to reasonable expectations for near future experiments. 

The axion-photon coupling constant ($\lambda_{true}$) used to generate signals depended on the background level under consideration, and $\lambda_{true} = 0$ was always used for sensitivity studies. Axion signals were generated according to Eq.~\ref{eq:compact} and event fluctuations were governed by Poisson statistics. The axion flux diminishes quickly at higher energy, and photon energy below 2 keV is too small to satisfy the Bragg condition. Therefore the energy region of interest (ROI) considered is between 2 and 8 keV. Without affecting the general results, a few simplifications are applied in the simulations. First, the detector $1\sigma$ energy resolution is assumed to be proportional to photon energy, $\sigma_E=0.04{\rm E}$, and the efficiency is assumed to be 100\%. The former is typical of past Ge-detector experiments, and the latter is also reasonable, given that energy thresholds below 2 keV are common.  

The experiment is assumed to be located at the Sanford Underground Research Facility (SURF)~\cite{Heise2014,Heise2015} with geographic coordinates (N44\degree 21' 10.75", W103\degree 45' 04.77"). This choice matches that of the MJD experiment~\cite{Abgrall2014, Xu2015}. The Sun's trajectory observed at SURF on an arbitrarily chosen day is obtained from a United States Naval Observatory (USNO) database~\cite{USNO}, and it is used repeatedly to produce many days of exposure. 

Due to fabrication processes, the [001] axis of a Ge crystal is normally aligned with the vertical axis, albeit a slight misalignment of few degrees is possible. A spherical coordinate system can be established for the laboratory room, with the crystal [001] axis being the z direction. To calculate the absolute solar angle, the only unknown parameter is the absolute azimuthal angle ($\phi$) of horizontal crystal axes, given that the Sun's position can always be calculated with high precision (about 0.2$\degree$ in the simulations). A Ge crystal is symmetrical about the [001] axis if rotated by 90$\degree$, so any one of the two horizontal axes can be used to define the absolute azimuthal angle, and the meaningful range of $\phi$ is -45$\degree$ to 45$\degree$. The precision of measurements for these angles impacts axion sensitivity. 

The simulated runs described here were analyzed under three assumed scenarios that represented different experimental implementations related to the absolute azimuthal angle. We refer to these as experimental scenarios 0, 1 and 2 (ES0, ES1, ES2). ES0 assumes that all angles are measured exactly. For ES1 all absolute angles of all crystals are measured to some precision. For ES2, the absolute angle is measured to some precision for only one crystal detector, but the relative crystal angles between that crystal and the others are measured with high precision. ES2 is the relative angle method we propose. The ideal situation of ES0 is not realizable in practice but it represents the best possible sensitivity, so both ES1 and ES2 are studied relative to ES0. 

For each combination of signal strength and background, 1000 experiments were simulated to form an ensemble to gain statistical significance. In every ensemble, hypothetical experimental data generated by Monte Carlo simulations were analyzed with the profile likelihood method~\cite{Rolke2005} to obtain confidence intervals (CI). If $\lambda_{true}=0$ in the simulations, there is no axion event and the ensemble averaged upper limit on $\lambda$ was calculated and referred to as the experimental sensitivity (of exclusion)\footnote{Here we follow Refs. ~\cite{Feldman1998, Gomez2011} and use ensemble average to define the sensitivity. Alternative definition involving ensemble median is also often used, \textit{e.g.} in Ref.~\cite{Cowan2011}.}. If $\lambda_{true}\ne 0$ in the simulations, the ensemble averaged width of the CI of $\lambda$ was calculated, which represents the experimental uncertainty, \textit{i.e.} how precisely the constant can be determined\footnote{This is different from experimental discovery potential.}. Both the experimental sensitivity and the CI were calculated at a 90\% confidence level (CL). We found the ensemble distribution of the test statistic sometimes deviates from the expected $\chi^2(1)$ distribution. Therefore, it is necessary to adjust the nominal critical value obtained from a $\chi^2(1)$ distribution to ensure proper coverage. This key step of numerically calculating the adjusted critical value is discussed in~\ref{app:stat}, along with other details of the statistical analysis.

\section{Experimental Scenario 0 with Known Angles}\label{sec:Known_Angle}

\begin{figure}[h]
\includegraphics[width=8cm]{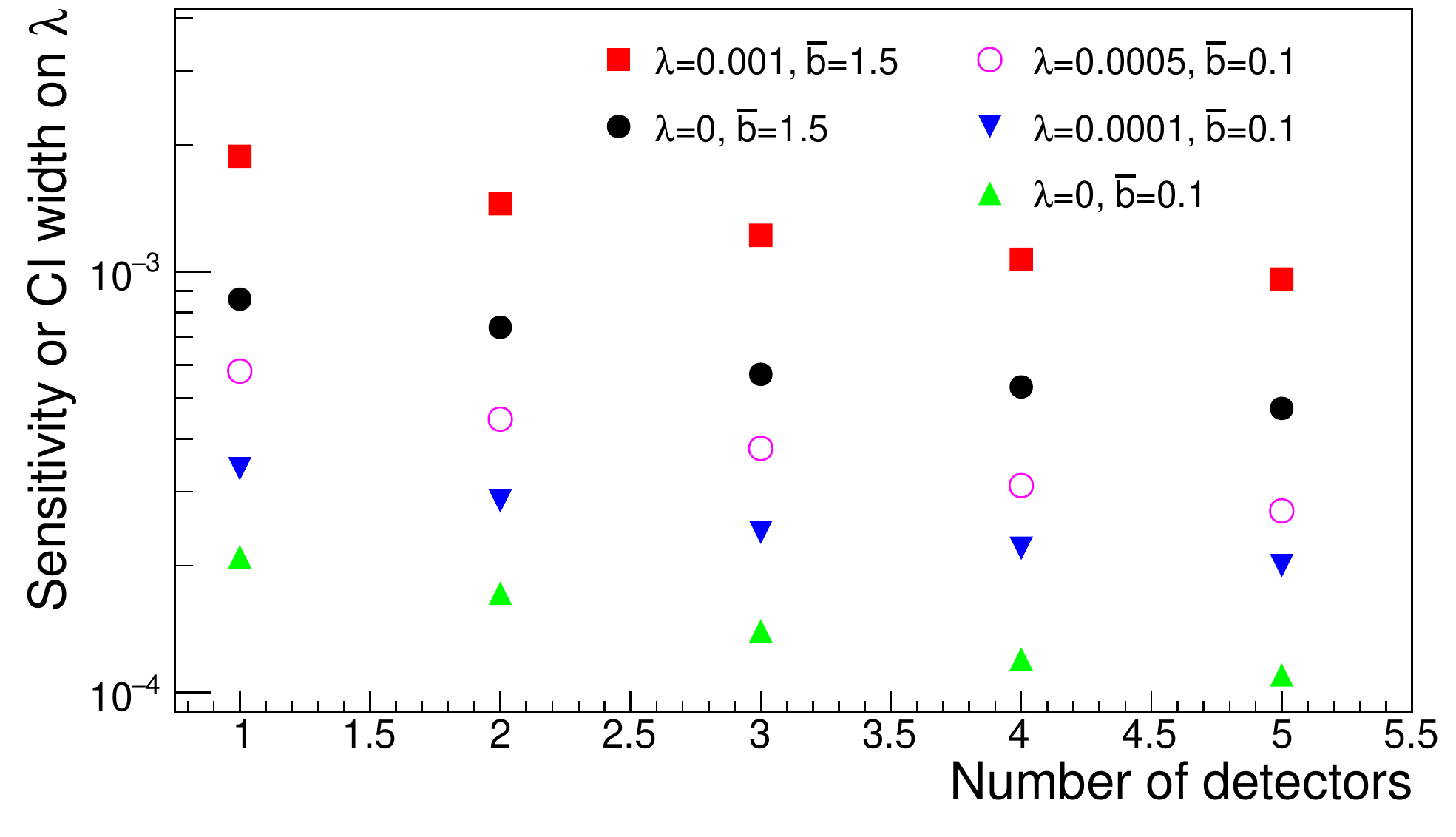}
\caption{Confidence interval width (sensitivity) for finite (zero) $\lambda_{true}$ in Ge-based experiments, extracted in ES0 with perfect knowledge of all detector angles. The background level ($\bar{b}$) unit is cts/(keV kg d). The exposure for each detector is 1000 (kg d).}
\label{fig:method0_all}
\end{figure}

\begin{figure}[h]
\includegraphics[width=8cm]{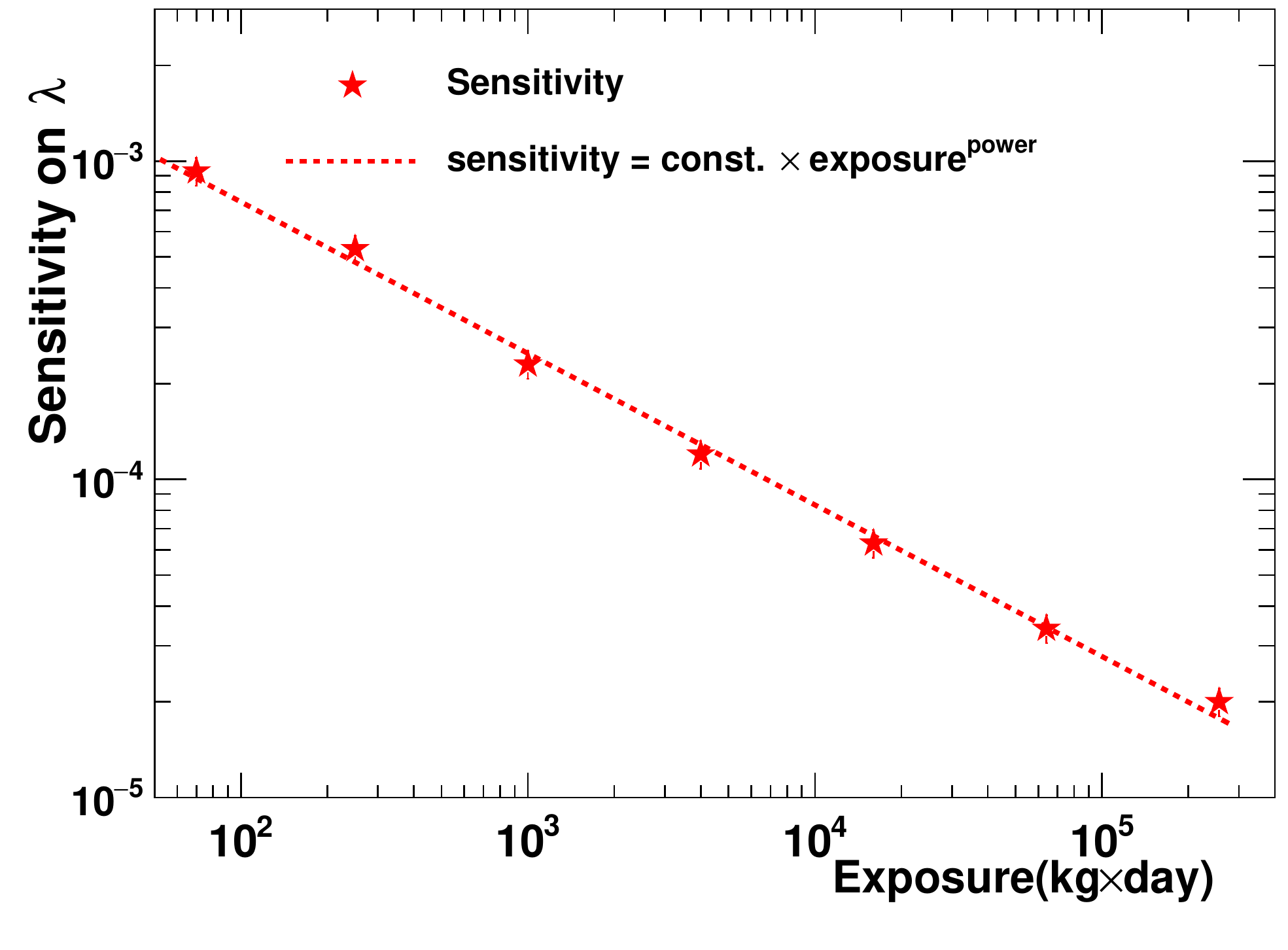}
\caption{Sensitivity on axion-photon coupling $\lambda$ extracted in ES0 as a function of total exposure. The assumed background level is $\bar{b}$=0.1 cts/(keV kg d), and energy resolution is $\sigma_E=0.04{\rm E}$. Vertical error bars presenting a 10\% relative statistical uncertainty due to the limited size of simulation ensembles are included to allow a fit. The fitted values are const. = $(6.7 \pm 0.9) \times10^{-3}$ and power = $-0.48 \pm 0.02$.}
\label{fig:exposure}
\end{figure}

\begin{figure}[h]
\includegraphics[width=8cm]{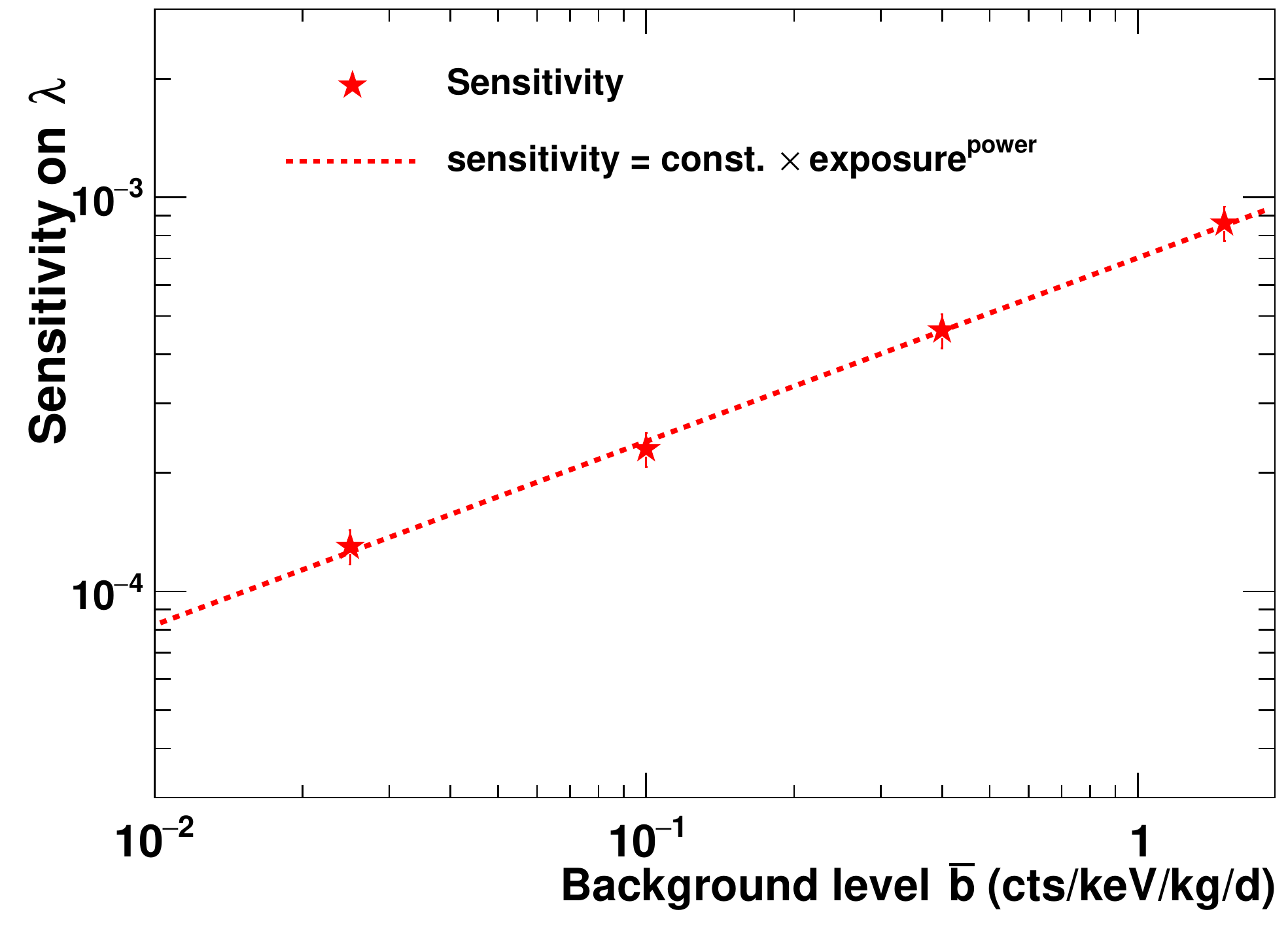}
\caption{Sensitivity on axion-photon coupling $\lambda$ extracted in ES0 as a function of background level. The assumed exposure is 1000 (kg d), and energy resolution is $\sigma_E=0.04{\rm E}$. Vertical error bars presenting a 10\% relative statistical uncertainty due to the limited size of simulation ensembles are included to allow a fit. The fitted values are const. = $(7.0 \pm 0.5) \times10^{-4}$ and power = $0.47\pm 0.03$.}
\label{fig:background}
\end{figure}

\begin{figure}[h]
\includegraphics[width=8cm]{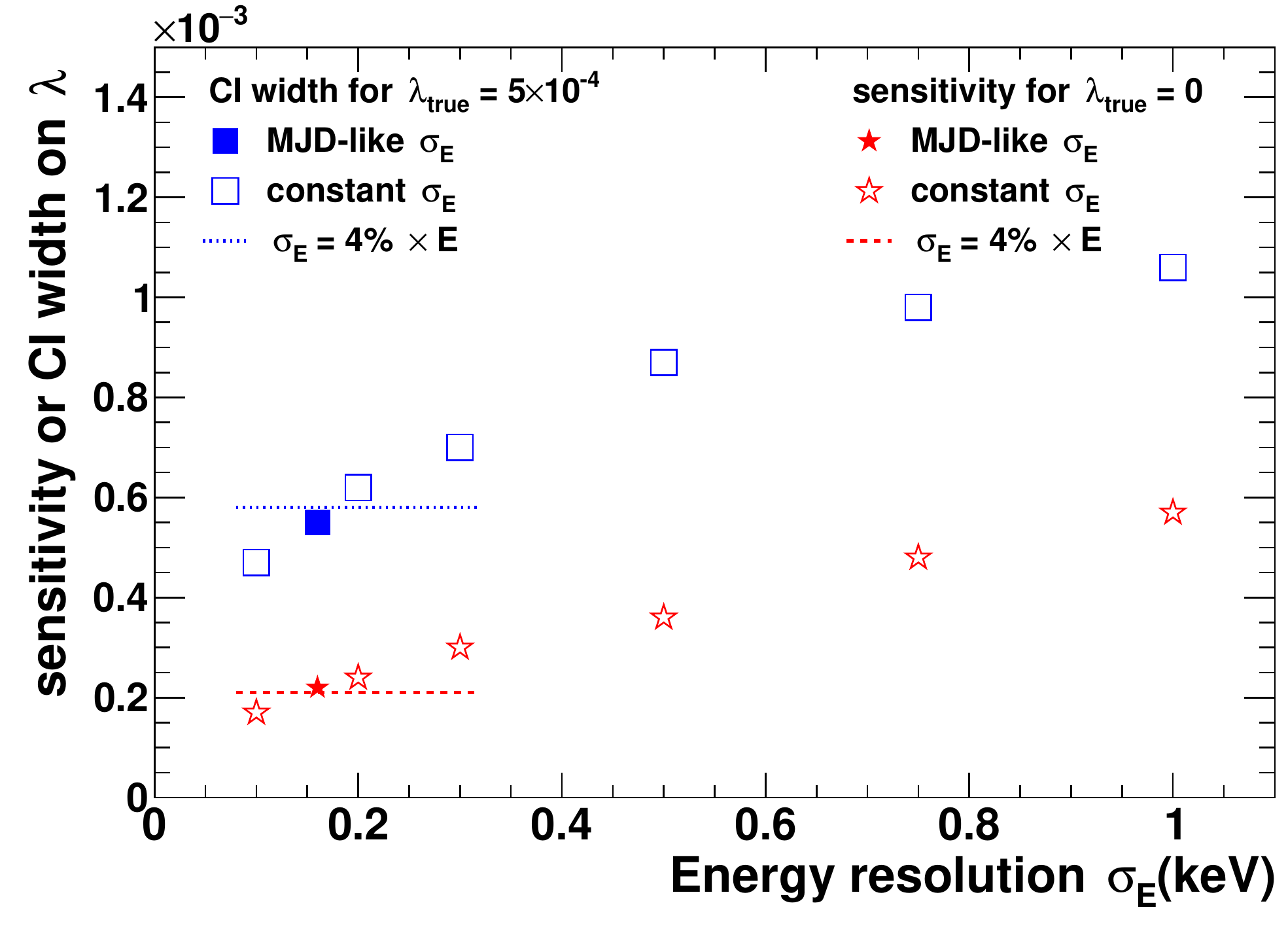}
\caption{Sensitivity or confidence interval width on axion-photon coupling $\lambda$ extracted in ES0 for different types of energy resolution. The assumed background level is $\bar{b}$=0.1 cts/(keV kg d), and the total exposure is 1000 (kg d). MJD-like energy resolution of Eq.~\ref{eq:MJD_sigmaE} is almost always 0.16 keV. The default energy resolution, $\sigma_E=0.04{\rm E}$, varies between 0.08 keV and 0.32 keV and the corresponding results are plotted as dotted and dashed lines.}
\label{fig:energy_resolution}
\end{figure}

The experimental scenario 0 (ES0) with perfect knowledge of all detector angles can achieve the best experimental sensitivity. To begin with some examples, Fig.~\ref{fig:method0_all} shows the ensemble averaged CI width and sensitivity on $\lambda$ in ES0 for several different combinations of background and signal strength for up to five detectors. The azimuthal angles of five detectors, $\phi^{(1)}_{true}=27.3{\degree}$, $\phi^{(2)}_{true}=-4.2{\degree}$,  $\phi^{(3)}_{true}=16.7{\degree}$, $\phi^{(4)}_{true}=35.0{\degree}$ and $\phi^{(5)}_{true}=-10.60{\degree}$, were arbitrarily chosen to generate axion signals when $\lambda_{true}\ne0$ is used in the simulations. Most of the data points here will sever as baselines in Sect.~\ref{sec:Relative_Angle} for ES1 and ES2 studies.

Due to the factor that each ensemble contains 1000 simulated experiments, the adjusted critical value for a 90\% CL was determined by the tail distribution made of 100 simulations, giving rise to a 10\% statistical uncertainty in general. This uncertainty was not included in any plots except in Figs.~\ref{fig:exposure} and ~\ref{fig:background}, where experimental sensitivity was quantitatively studied for various total exposures and background levels in ES0 with $\sigma_E=0.04{\rm E}$. A linear fit to the log-log plot of Fig.~\ref{fig:exposure} reveals that the improvement of sensitivity is consistent with $\sqrt{\rm exposure}$ for up to 2.56$\times10^{5}$ (kg d) in our simulations. This 2.56$\times10^{5}$ (kg d) exposure is predicted to provide a sensitivity of roughly 2$\times10^{-5}$ on $\lambda$ or 7$\times10^{-10}~\rm GeV^{-1}$ on g$_{a\gamma\gamma}$ at 90\% CL in ES0 with the specific experimental parameters in Fig~\ref{fig:exposure}. 

Similarly, the worsening of sensitivity is consistent with $\sqrt{\rm background}$ for ${\bar{b}}$ between 0.025 and 1.5 cts/(keV kg d), as shown in Fig.~\ref{fig:background}. These square root relationships are not surprising, because many statistical quantities are often strongly correlated to the square root of sample size, which is essentially the product of background rate and exposure in simulated datasets without signals. However, this square root behavior may be affected by the physical boundary at zero when the sensitivity keeps improving with larger exposure or lower background. 
 
The default energy resolution in our studies, $\sigma_E=0.04{\rm E}$, translates to a value between 0.08 and 0.32 keV in the 2 to 8 keV energy region of interest for solar axions. Recently, the MJD experiment has achieved an excellent energy resolution based on purely ionization energy loss, as Eq.~\ref{eq:MJD_sigmaE} ~\cite{Abgrall2016}. For energy between 2 and 8 keV, the second term in the square root of Eq.~\ref{eq:MJD_sigmaE} is less than 10\% of the first term, so the energy resolution of MJD is essentially a constant of 0.16 keV.
\begin{equation}
\label{eq:MJD_sigmaE}
\sigma_E^{\rm MJD}=\sqrt{(0.16~ \rm keV)^2 + 0.11\times2.96~ \rm eV\times\rm E}
\end{equation}

Experimental sensitivity and precision (confidence interval width) corresponding to a MJD-like energy resolution (Eq.~\ref{eq:MJD_sigmaE}) were studied, as shown in Fig.~\ref{fig:energy_resolution}. In addition, constant energy resolution between 2 and 8 keV was also studied assuming a span of values. As expected, experimental sensitivity and precision degrade with worse energy resolution. Results using the default energy resolution $\sigma_E=0.04{\rm E}$ are also plotted in Fig.~\ref{fig:energy_resolution} for comparision, and they are almost identical to those using the MJD-like energy resolution for both zero and finite $\lambda$, suggesting that both types of energy resolution are comparably excellent. 

\section{Precision of Absolute Angle Measurement}\label{sec:angle_precision} 
While exact knowledge of all absolute angles allows the best possible experimental sensitivity in Sect.~\ref{sec:Known_Angle}, it is experimentally impractical. Hence, it is important to quantify a reasonably and practical precision for absolute angle measurements. Moreover, as demonstrated by the Monte Carlo studies in~\ref{app:stat}, the precision of the absolute angle measurement affects the critical values required to properly set a confidence interval in the profile likelihood method. It is therefore beneficial to know the maximum angular uncertainty for which nominal critical values based on $\chi^2(1)$ still can be applied without causing under-coverage. 

The [001] axes of Ge crystals are assumed to be exactly vertical, although a misalignment is possible in reality~\cite{Ahmed2009, Bruyneel2006a} and may have potential adversary impacts for future experiments. The absolute angle measurements mainly concern the azimuthal angles ($\phi$). Monte Carlo runs were generated with a span of experimental precision on the absolute azimuthal angles so that sensitivity as a function of angle uncertainty can be studied.  A single detector with 1000 (kg d) exposure was simulated for both $~\bar{b}=0.1$~and 1.5 cts/(keV kg d) with $\lambda_{true}=0$. 

\begin{figure}[h]
\includegraphics[width=0.5\textwidth]{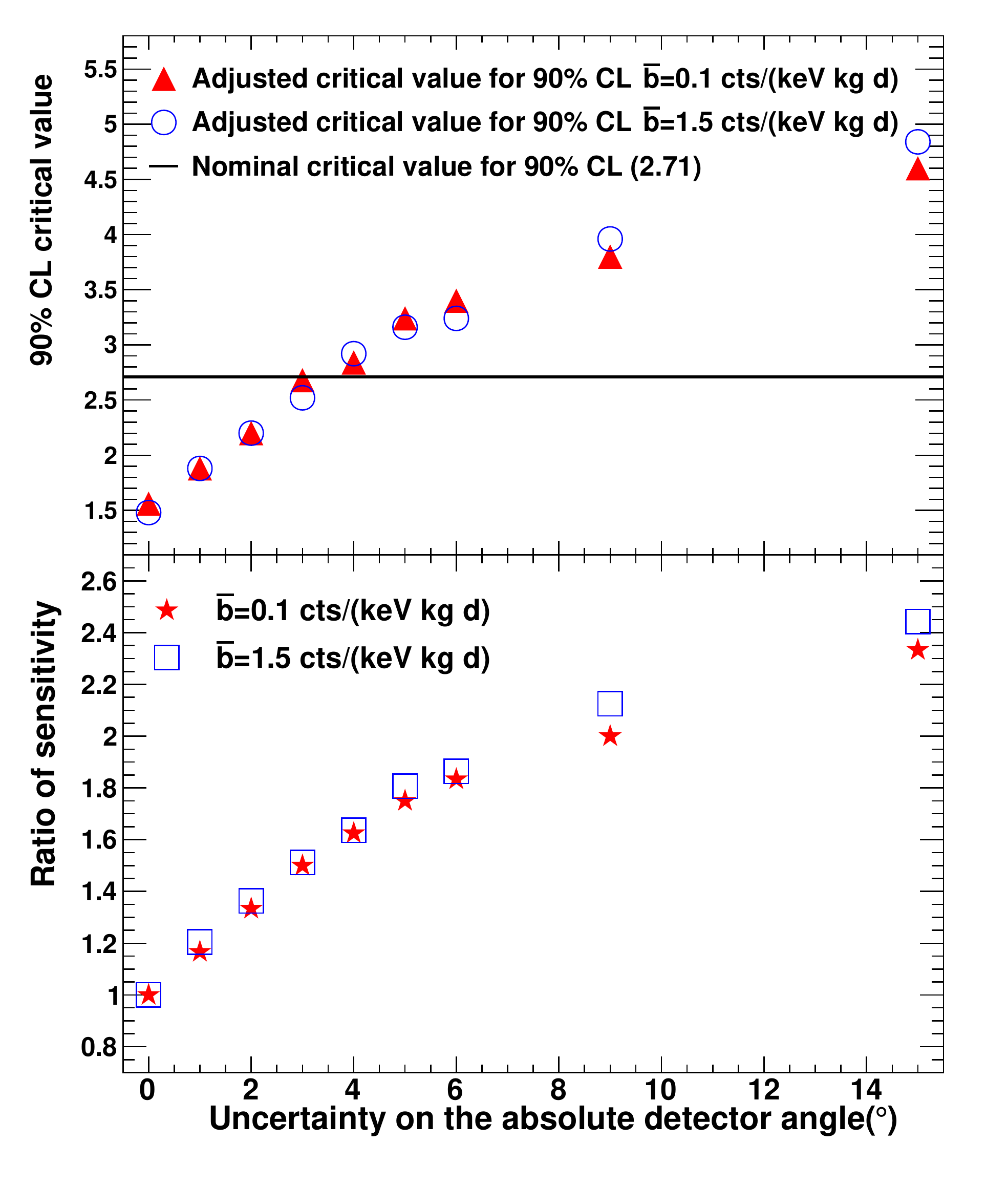} 
\caption{The adjusted critical value required to obtain a confidence interval at a 90\% CL (upper panel) and the relative sensitivity on axion-photon coupling compared to the case of zero angular uncertainty (lower panel), as functions of uncertainty on the absolute azimuthal angle. The leftmost points at zero angular uncertainty correspond to ES0. The horizontal line in the upper panel indicates the nominal critical value of 2.71 for a 90\% CL, obtained directly from a $\chi^2(1)$ distribution.}
\label{fig:angle}
\end{figure}

The upper panel of Fig.~\ref{fig:angle} shows how the critical value required for a 90\% confidence interval grows with angular uncertainty. For an angular uncertainty of 4$\degree$ or larger, the nominal critical value of 2.71 from $\chi^2(1)$ is not sufficient to reach 90\% CL in the profile likelihood method any more. For smaller angular uncertainty, an experiment does not necessarily need to perform adjustment to the nominal critical values, if over-coverage is not a concern. 

Poorer angular measurement also expectedly causes worse experimental sensitivity. This can be seen in the lower panel of Fig.~\ref{fig:angle}, in which all results are normalized to the ideal case with zero angular uncertainty. At small angular uncertainty of a few degrees, the experimental sensitivity worsens roughly linearly with the angular uncertainty. For example, for uncertainties of 2$\degree$ and 4$\degree$, the experimental sensitivities are about 35\% and 60\% worse than ES0 with exact angular knowledge, respectively. 

With a flat background model, little dependence on the actual background level is observed in either panel of Fig.~\ref{fig:angle}. Therefore, it appears a measurement of the absolute azimuthal angle to within~$2\degree$ to~$4\degree$ is acceptable and does not require Monte Carlo-based statistical studies. It should be noted that because of the Bragg condition, poor energy resolution will diminish the advantage of a precise angular determination. The energy resolution used in Fig.~\ref{fig:angle} is $\sigma_E=0.04{\rm E}$. 

\section{Relative Angle Method}\label{sec:Relative_Angle}
Measurements of absolute detector angles with respect to the Sun often require comprehensive surveys on the orientations of research facility, laboratory room, experimental apparatus and specific installation of each crystal. Assuming just a few degrees of angular uncertainty introduced in each step, the total uncertainty can add up quickly. Even a~$2\degree$ to~$4\degree$ precision as suggested in Sect.~\ref{sec:angle_precision} could be challenging to achieve. 

In contrast, relative angles between detectors can be measured \textit{in situ} without involving complicated surveys. Ref.~\cite{Bruyneel2006a} extracted crystal orientation in a local coordinate system for a segmented Ge detector by measuring and model-fitting the anisotropy in the charge carrier drift velocity with respect to the crystal axes. Based on the variation of best fit values reported in Tables 1 and 2 of~\cite{Bruyneel2006a}, we conclude that a precision of about $\pm1\degree$ can be achieved for the local crystal orientation, given enough statistics. For unsegmented Ge detectors, the same anisotropy can be probed with the help of photons in coincidence, such as the 81 keV and 356 keV photons of $^{133}$Ba. The anisotropy will be imprinted in the time difference between lower energy photon signals in a Ge detector and higher energy photon signals from a fast radiation detector, \textit{e.g.} scintillators coupled with Photomultiplier Tubes (PMT). Once a local crystal orientation is measured accurately, the relative crystal orientation between any two detectors can be calculated out simply as the difference between the two local orientations.

The same simulation framework as before were used here, and there were $n_D$ angles for $n_D$ detectors. The computational requirements were extensive, therefore only simulations with five or fewer detectors were considered. The detectors were taken to be individually 1~kg in mass, and a total of 1000 (kg d) was simulated for each detector in every hypothetical experimental run. We considered poor precision of $\pm7.5\degree$ for the absolute angle measurements and good precision of $\pm2\degree$ for relative angle measurements. These uncertainties are mainly systematic in nature, and thus they were assumed to follow uniform distributions and randomized accordingly in the simulations. 

While experimental scenario 1 (ES1) uses only absolute angles, experimental scenario 2 (ES2) utilizes also relative angles with respect to the $1^{st}$ detector. For every detector in ES1 and the $1^{st}$ detector in ES2, the hypothetically measured absolute angle ($\phi^{(j)}_{mea}$) was randomly generated from a uniform distribution $U(\phi^{(j)}_{true}-7.5{\degree}, \phi^{(j)}_{true}+7.5{\degree})$, where $j$ is the detector index and $\phi^{(j)}_{true}$ is the true value. In ES2, the hypothetically measured relative angle ($\Delta\phi^{(j)}_{mea}, j=2, 3, 4, 5$) was generated in a similar fashion around the true relative angle $\phi^{(j)}_{true}-\phi^{(1)}_{true}$. When analyzing the simulated data sets, the allowed range for absolute and relative angles were constrained around the hypothetically measured values within the corresponding uncertainties. Both ES1 and ES2 involve practical scenarios of measured angular information. In contrast, all absolute angles were assumed to be exactly known, $\phi^{(j)}_{mea}=\phi^{(j)}_{true}$, for ES0. 

\begin{figure}
\centering

	\begin{subfigure}[b]{0.5\textwidth}
	\includegraphics[width=1\textwidth]{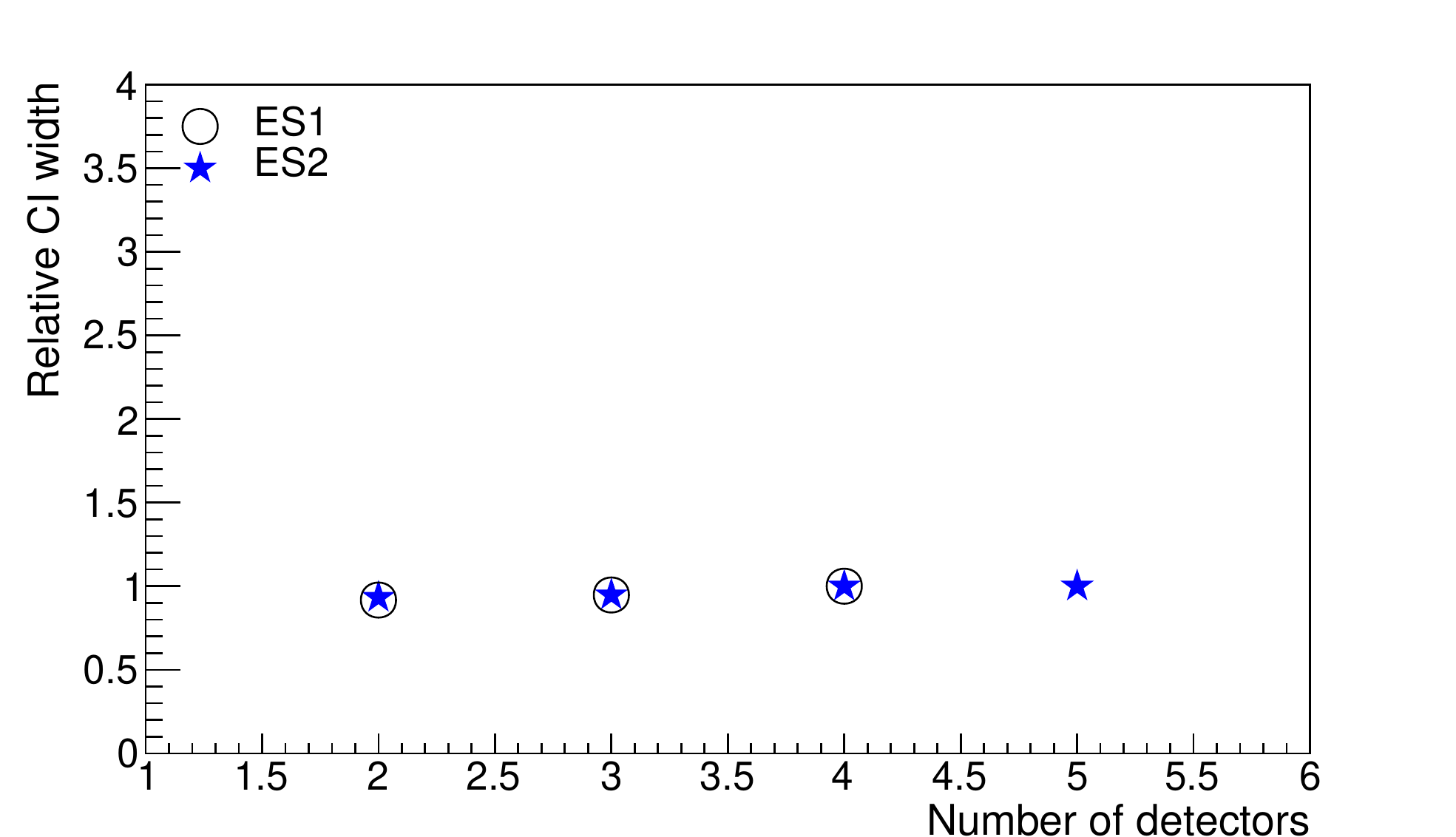}
	\caption{~$\lambda_{true}=0.0005,~\bar{b}=0.1$~cts/(keV kg d)}
	\label{fig:L0.0005_B0.60}
	\end{subfigure}

	\begin{subfigure}[b]{0.5\textwidth}
	\includegraphics[width=1\textwidth]{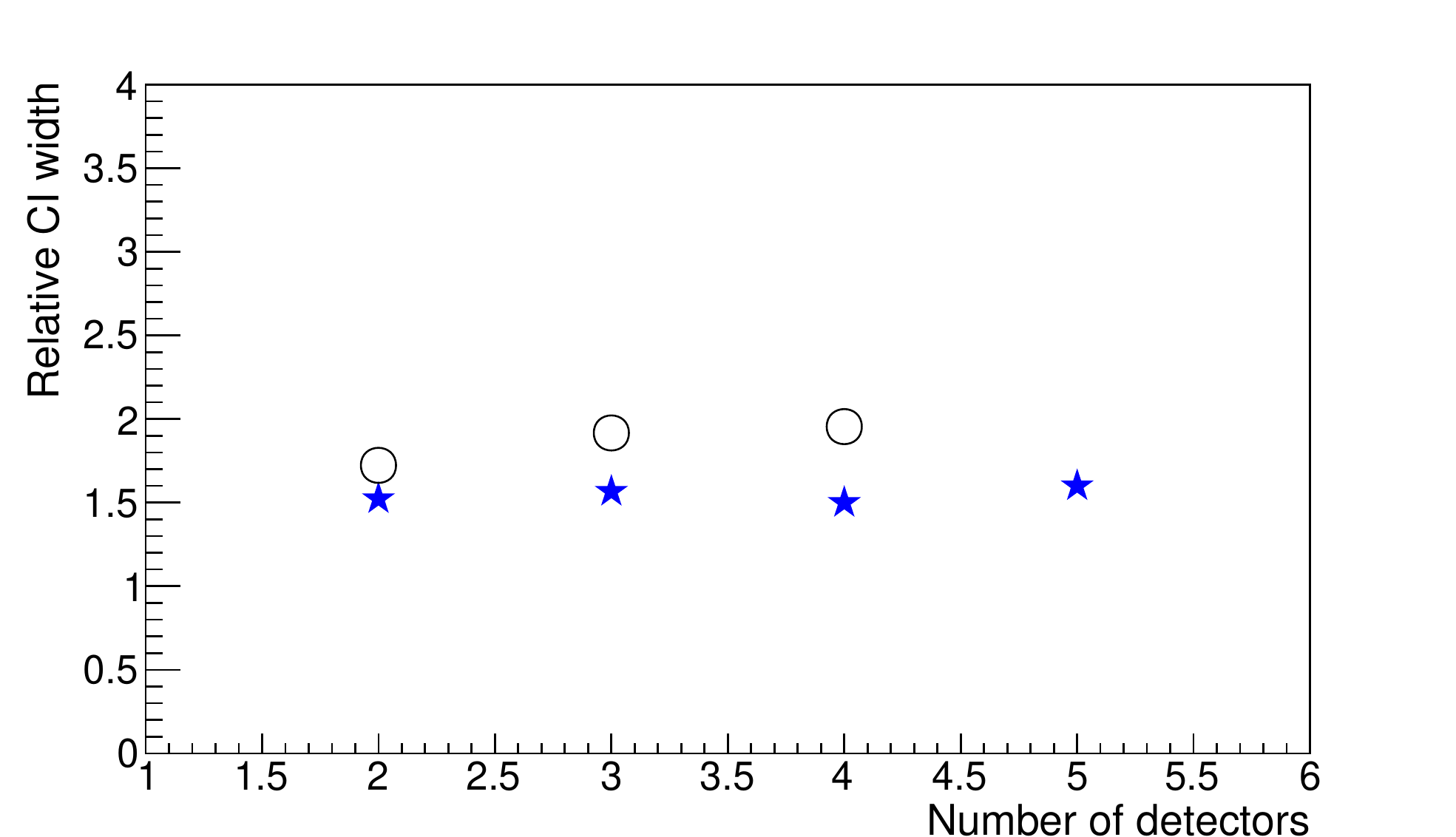}
	\caption{~$\lambda_{true}=0.0001,~\bar{b}=0.1$~cts/(keV kg d)}
	\label{fig:L0.0001_B0.60}
	\end{subfigure}

	\begin{subfigure}[b]{0.5\textwidth}
	\includegraphics[width=1\textwidth]{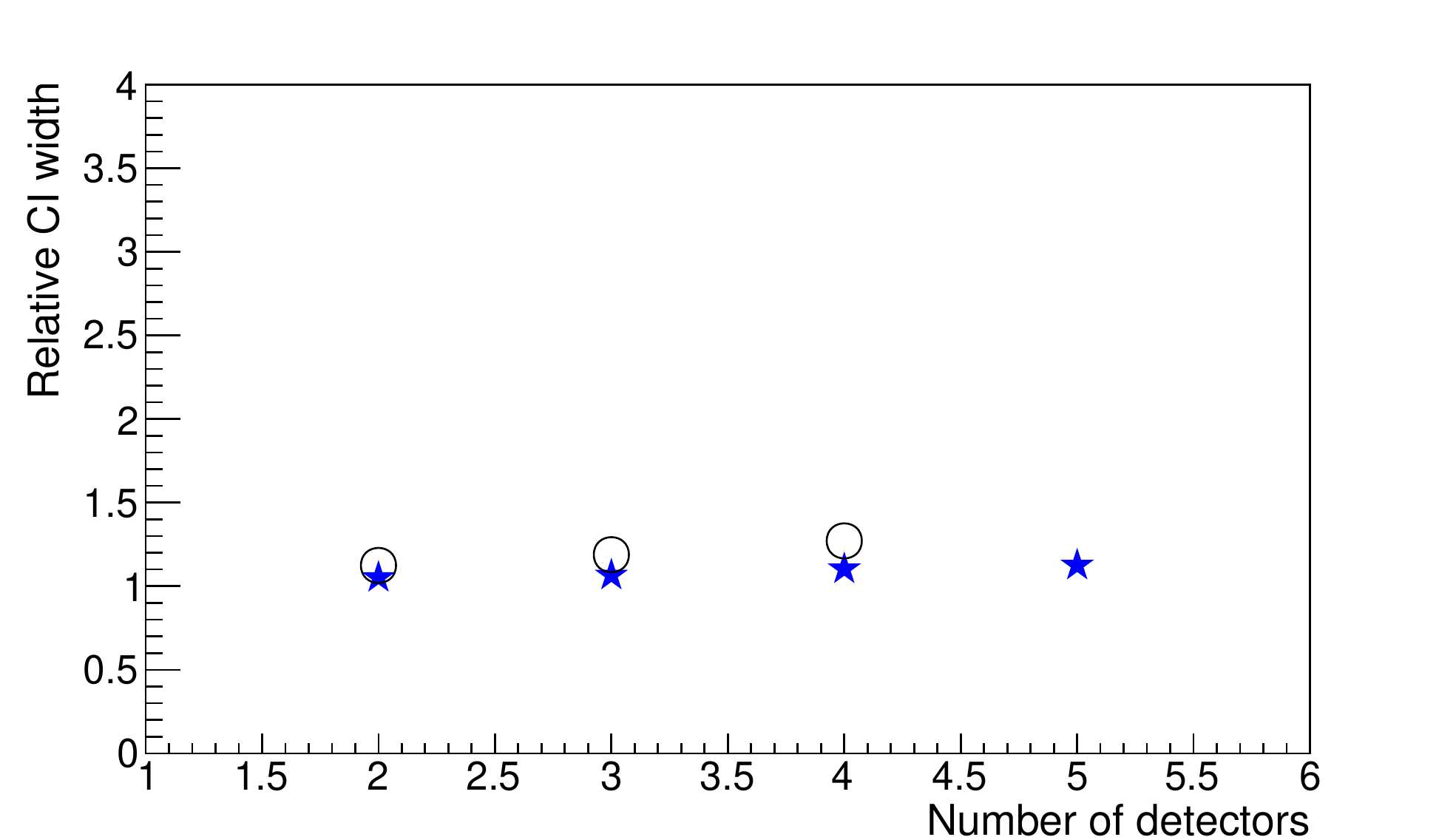}
	\caption{~$\lambda_{true}=0.001,~\bar{b}=1.5$~cts/(keV kg d)}
	\label{fig:L0.0010_B9.00}
	\end{subfigure}
\caption{ES1 (circles) and ES2 (stars) confidence interval width relative to the ES0 results, which are shown in Fig.~\ref{fig:method0_all}. Only ES2 is considered for 5 detectors, due to higher computational requirements of ES1. A smaller value of relative CI width indicates a better performing experiment.}
\label{fig:L_nonzero}
\end{figure}
\begin{figure}[h]
\centering
	\begin{subfigure}[b]{0.5\textwidth}
	\includegraphics[width=1\textwidth]{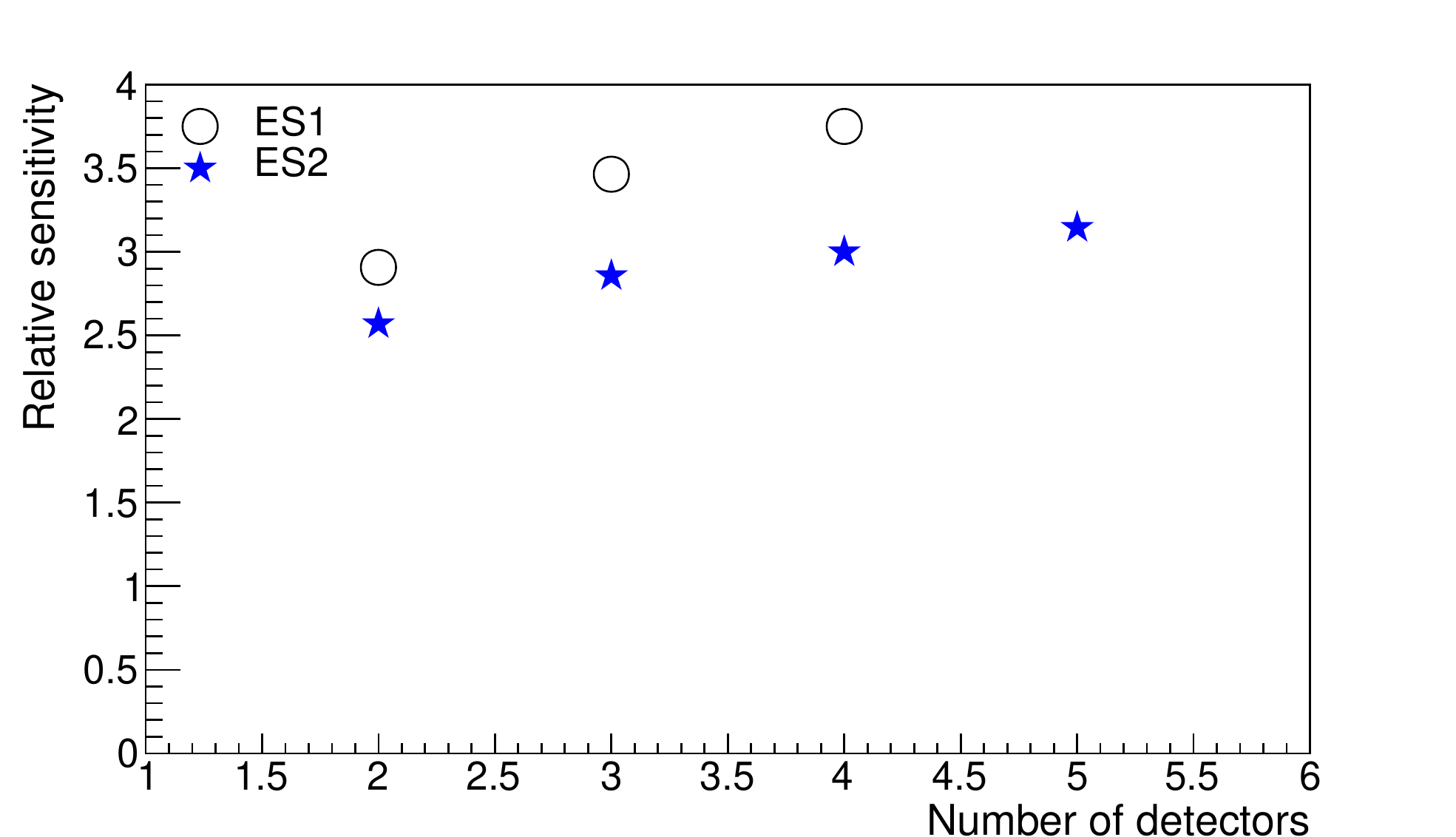}
	\caption{~$\lambda_{true}=0,~\bar{b}=0.1$~cts/(keV kg d)}
	\label{fig:L0.0000_B0.60}
	\end{subfigure}
	
	\begin{subfigure}[b]{0.5\textwidth}
	\includegraphics[width=1\textwidth]{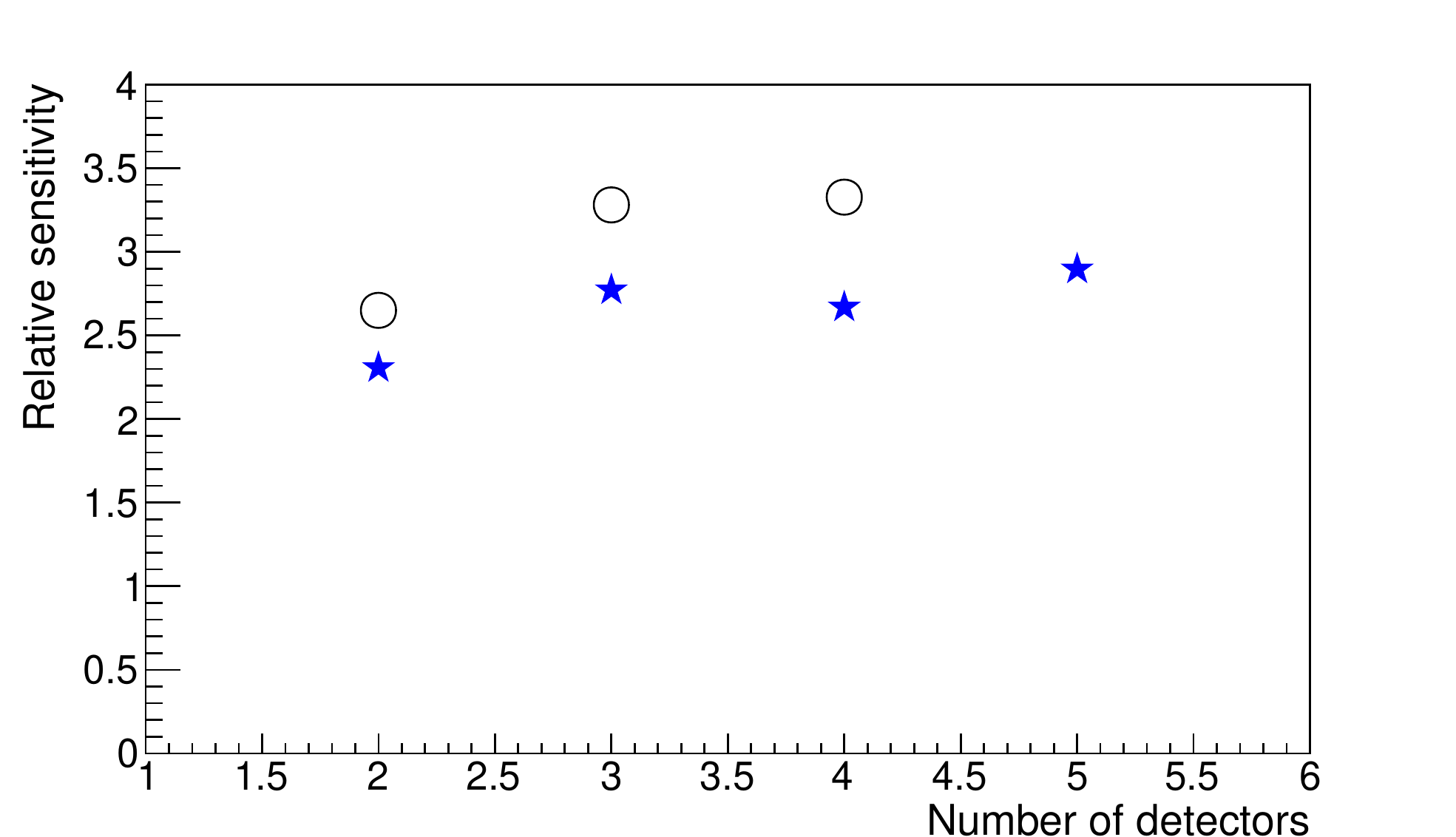}
	\caption{~$\lambda_{true}=0,~\bar{b}=1.5$~cts/(keV kg d)}
	\label{fig:L0.0000_B9.00}
	\end{subfigure}
\caption{ES1 (circles) and ES2 (stars) sensitivity relative to the ES0 results, which are shown in Fig.~\ref{fig:method0_all}. Only ES2 is considered for 5 detectors. The signal-to-background ratio is alway zero. A lower relative sensitivity indicates a better performing experiment.}
\label{fig:L_zero}
\end{figure}

For easy comparison, the results of ES1 and ES2 are expressed as ratios to ES0. For finite $\lambda_{true}$, confidence interval width relative to ES0 is calculated as $\frac{\hat\lambda_{up}^{(i)}-\hat\lambda_{lo}^{(i)}}{\hat\lambda_{up}^{(0)}-\hat\lambda_{lo}^{(0)}}$, where $\hat\lambda_{up}$ and $\hat\lambda_{lo}$ are the upper and lower boundary of the confidence interval estimated in each simulation, respectively. The superscript is the experimental scenario index, so $\hat\lambda^{(0)}$ is for ES0 and $i=1, 2$. For $\lambda_{true}=0$, experimental sensitivity relative to ES0 is calculated. For the same ensemble of simulations, a larger value of either the relative CI width or the relative sensitivity indicates a less competitive experimental method. Figures~\ref{fig:L_nonzero} and~\ref{fig:L_zero} show the comparisons between ES1 and ES2.

Generally speaking, utilizing the more precisely measured relative angles, ES2 has a significant improvement over ES1. Both the signal-to-background ratio and the absolute signal strength play roles in the comparison of these two experimental scenarios. A comparison between Fig.~\ref{fig:L0.0005_B0.60} and Fig.~\ref{fig:L0.0010_B9.00} suggests that the improvement of ES2 over ES1 is more prominent in the case of lower signal-to-background ratio. In the extreme case of Fig.~\ref{fig:L0.0005_B0.60}, the signal-to-background ratio is high enough that the axion signals stand out in the data clearly and the detector angles can be estimated out in the profile likelihood analysis. In this case, both ES1 and ES2 have similar performance to ES0. In contrast, although Fig.~\ref{fig:L0.0010_B9.00} has twice the number of signal events, the much larger background in the data prohibits the distinct signal pattern from being clearly recognized. Therefore, background reduction is critical.

When the axion-photon coupling approaches zero, the improvement of ES2 over ES1 also becomes more significant. For example, although Fig.~\ref{fig:L0.0001_B0.60} ($\lambda_{true}=0.0001,~\bar{b}=0.1$~cts/(keV kg d)) has a bit higher signal-to-background ratio than Fig.~\ref{fig:L0.0010_B9.00}, it also has a factor of 10 fewer signal counts. As a result, the advantage of ES2 is more obvious in Fig.~\ref{fig:L0.0001_B0.60}. 

In the limiting case where there are no axion signals, the advantage of ES2 over ES1 becomes the most significant, as shown in Fig. ~\ref{fig:L_zero}. 
This is supported by simple statistical agreements. As demonstrated in ~\ref{app:stat}, the $\pm7.5\degree$ uncertainty is large enough that detector absolute angles are effectively unknown. Therefore, almost $n_D+1$ parameters are unknown in ES1. Utilizing good knowledge of relative angles in ES2, detector absolute angles are tightly constrained against the first detector, significantly reducing the number of unknown parameters presented in the problem. This is the root of ES2 advantages.

Based on the data points in Figs.~\ref{fig:L_nonzero} and~\ref{fig:L_zero}, the improvement of ES2 relative to ES1 grows with $n_D$. This is expected as the advantage of the good knowledge of relative angles over the poor knowledge of absolute angles grows with more detectors. This improvement reaches as high as approximately 30\%, only limited by the number of detectors studied here, representing essentially a 30\% better experimental sensitivity in ES2 using the relative angle method. The correlation between energy and angle due to the Bragg condition means that poor energy resolution will also reduce the advantage of a precise relative angular measurement, quenching the improvement of ES2 over ES1.

\section{Averaging Over Detector Angles}\label{sec:averaging}
Experiments become more challenging with more detectors, in part due to increasing effort required to measure all crystal angles, absolute or relative. However, a different technique to correlate multiple detectors becomes possible with a large number of detectors orientated randomly. In this case, the axion signal probability function can be averaged over absolute azimuthal angles. This method has been applied previously, in particular by DAMA~\cite{Bernabei2001}, where an average over nine NaI(Tl) crystal scintillators was used. 

Modern solid state detectors are often bulk produced and thus these detectors are often similar to each other in both appearance and performance. For example, the natural Ge detectors in the {\sc Majorana Demonstrator} (MJD) experiment~\cite{Abgrall2014, Xu2015} are very similar in every aspect, including mass. For an experiment with a very large number of identical detectors ($n_D>>1$) oriented with no regard to crystal axis, the actual distribution of detector azimuthal angles will closely resemble a uniform distribution. In this case, the distribution of axion signals summed over all $n_D$ detectors of mass $M_D$ approximates the same distribution as that averaged over angle for a hypothetical single detector with a mass of $n_D \times M_D$.

\begin{figure}[h]
\includegraphics[width=0.5\textwidth]{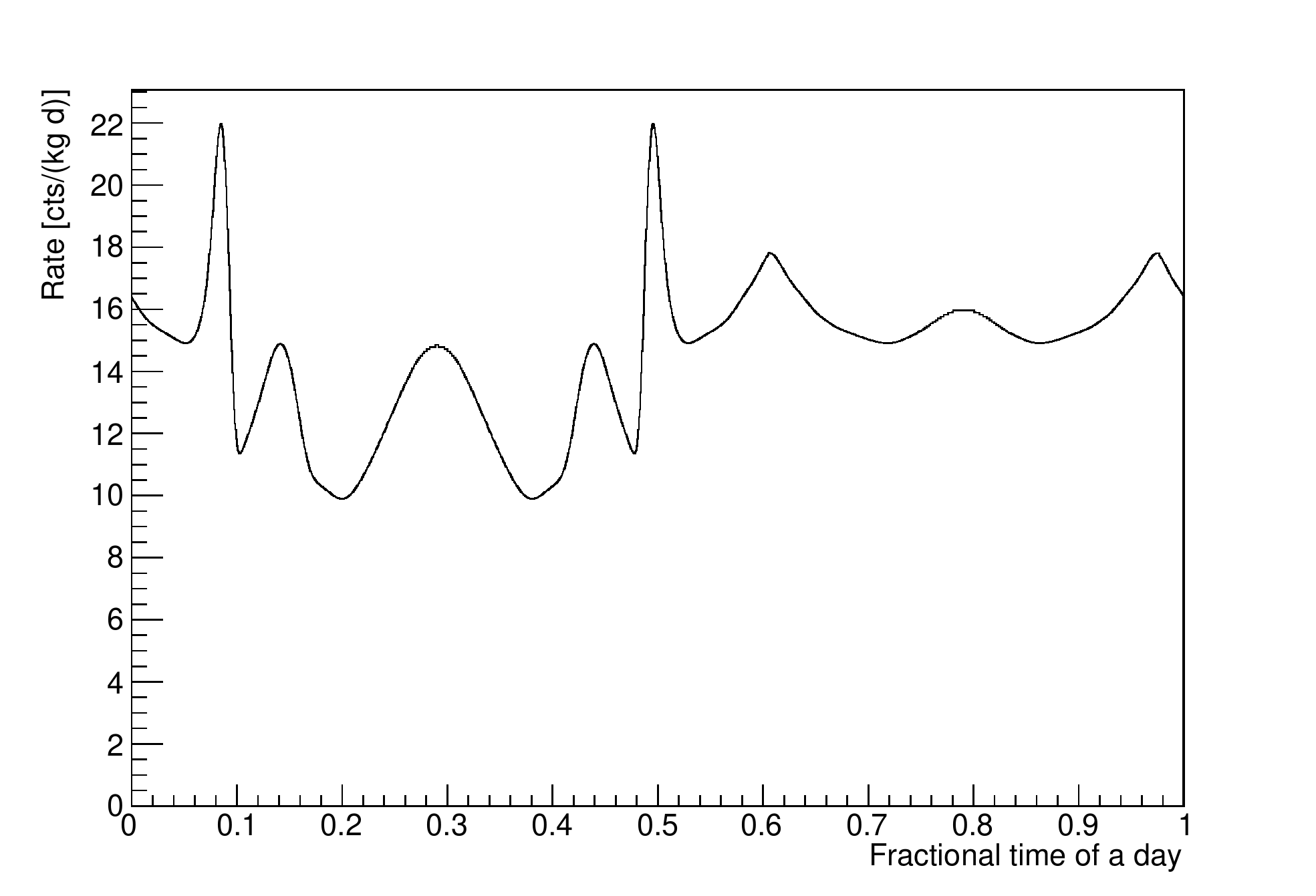}
\caption{Azimuthal angle-averaged axion-induced photon signal rate, \textit{i.e.} the right side of Eq.~\ref{eq:average}, with respect to time of a day with photon energy between 4.0-4.5keV for $g_{a\gamma\gamma}=10^{-8}$ GeV$^{-1}$ ($\lambda=1$) in a HPGe detector located at Lead, SD.}
\label{fig:Energy_averaged_angle}
\end{figure}

The actual signal probability can be approximated by the signal probability averaged over all possible azimuthal angles,
\begin{equation}
\frac{\sum_{j=1}^{n_D}\dot R^{(j)}(\phi^{(j)})}{n_D}\rightarrow \frac{\int^{\pi/4}_{-\pi/4} \dot R(\phi)d\phi}{\pi/2}, ~\rm{if}~ n_D\rightarrow \infty~,
\label{eq:average}
\end{equation}
where $\dot R(\phi)$ is just Eq.~\ref{eq:compact} with axis angle explicitly noted as a parameter. An example of axion signal pattern from the right side of Eq.~\ref{eq:average} is shown in~Fig.~\ref{fig:Energy_averaged_angle}.

Although $n_D\rightarrow \infty$ is impractical, experiments with dozens or more of identical Ge detectors, or other solid-state detectors, are ongoing and future experiments with hundreds of identical detectors are currently being considered. The number of detectors required to ensure that the left side of Eq.~\ref{eq:average} well approximates the right side is the critical question studied here.

We simulated different experimental configurations with 1, 5, 15, 50 and 150 identical Ge detectors. The total mass of all detectors in each configuration was 100~kg and the live time was 1000 d, so that the total exposure in each simulation was $10^5$~(kg d), regardless of the number of detectors deployed. Each configuration was simulated 1000 times to produce an ensemble. The detector angles were randomly generated in each simulation and the corresponding axion signals were simulated according to Eq.~\ref{eq:compact}, with $\lambda_{true}=0$ and 0.0005. Only $\bar{b}=0.1$~cts/(keV kg d) was simulated and it is the sole nuisance parameter. The simulated data sets were analyzed using the method detailed in~\ref{app:stat}, except that the constructed likelihood function uses the right side of Eq.~\ref{eq:average} to describe axion signals in all detectors together\footnote{The statistical consideration doesn't involve a joint likelihood function of individual detectors any more, but a single likelihood function for the entire experiment. Individual detector information is ignored to accommodate unknown individual detector angles}.

Independently, an experiment with $n_D\rightarrow \infty$ was simulated in an indirect but equivalent way. As argued above, Eq.~\ref{eq:average} holds exactly in this case, and therefore the axion signals were first simulated and then analyzed both according to the right side of Eq.~\ref{eq:average}. This represents the optimal sensitivity for the angle averaging method. As before, ES0 with perfect angular information was also studied to obtain a baseline for the same total exposure, where the angle-dependent axion signal probability of individual detectors was used to construct the joint likelihood function. 

\begin{figure}[h]
\includegraphics[width=0.5\textwidth]{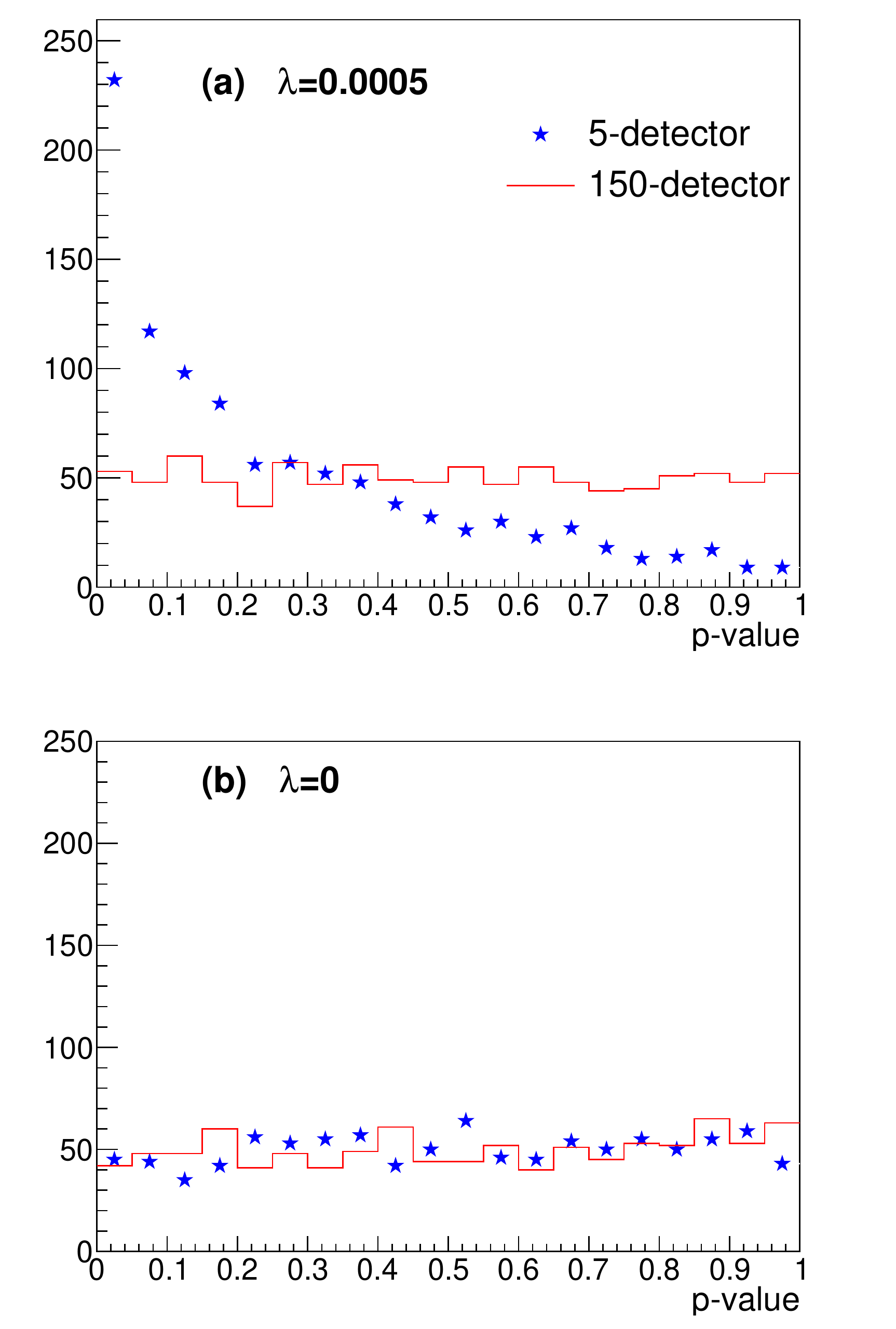}
\caption{ P-value distribution for GoF tests between the simulation data sets and the models using the angle-averaging method for 5-detector (blue star) and 150-detector (red histogram) experiments. Panel (a)~$\lambda_{true}$=0.0005 and (b)~$\lambda_{true}$=0, with$~\bar{b}=0.1$~cts/(keV kg d). }
\label{fig:gof}
\end{figure}

A $\chi^2$ goodness of fit (GoF) test~\cite{Olive2014} was carried out to quantify the consistency between the simulation data sets and the best-fit models in the likelihood analysis. Then the corresponding p-value for each GoF test result was calculated with ROOT~\cite{Brun97}. Fig.~\ref{fig:gof} compares the p-values of the models using the angle-averaging method for 5-detector and 150-detector configurations. For non-zero axion-photon coupling in panel (a), there is a concentration of small p-values in the 5-detector experiment, indicating that the model doesn't agree with the data. In contrast, the p-values for the 150-detector experiment uniformly distribute between 0 and 1, indicating a good consistency. Therefore, Eq.~\ref{eq:average} is a good approximation to model any single experiment with 150 detectors but not for one with only 5 detectors. 

However, for zero $\lambda_{true}$ in panel (b) of Fig.~\ref{fig:gof}, both 5-detector and 150-detector experiments have uniformly distributed p-values, indicating the averaged axion signal probability can describe the data in both cases. This is not surprising given that there is no axion signal at all in the simulated data sets. Any model of signal could be combined with a good background model to describe the data well within the statistical uncertainties, as long as the background rate dominates in the fitted model.  Furthermore, even if Eq.~\ref{eq:average} doesn't hold and the angle-averaged axion pattern, \textit{e.g.} Fig.~\ref{fig:Energy_averaged_angle}, is not as distinct as the actual signal pattern, \textit{e.g.} Fig.~\ref{fig:Energy}, it is far from a featureless background. As a result, it still has some discriminating power against the background. 

\begin{figure}[h]
\includegraphics[width=0.5\textwidth]{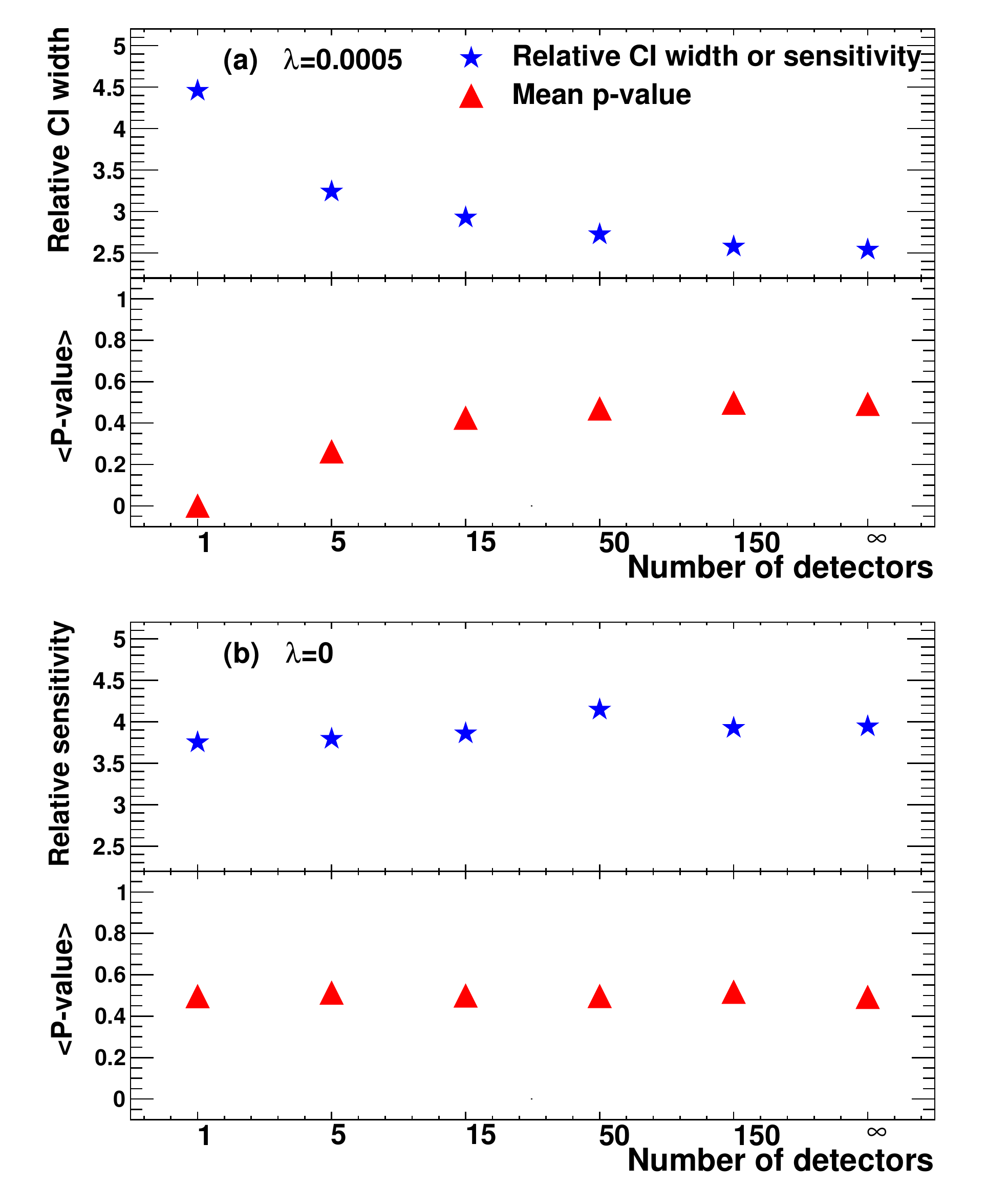}
\caption{Panel (a): Relative confidence interval widths for~$\lambda_{true}=0.0005,~\bar{b}=0.1$~cts/(keV kg d) and $\sigma_{E}=0.04\rm E$ from models using the angle-averaging method, shown as blue stars in the upper half of the plot. The corresponding mean p-values are plotted as red triangles in the lower half. The optimal case of $n_D\rightarrow \infty$ is shown as the rightmost points. Panel (b): Relative sensitivity and mean p-values for ~$\lambda_{true}=0$.}
\label{fig:plot_all_multiple}
\end{figure}

Fig.~\ref{fig:plot_all_multiple} shows the relative widths of confidence intervals and sensitivities at a 90\% CL. As before, the results from the averaging method are divided by the corresponding values in ES0. Experiments with $n_D\rightarrow \infty$ are shown as the rightmost data points. The mean p-value in each ensemble is also shown, and a mean p-value that is much less than 0.5 indicates the model doesn't describe the data.

In the case of a large signal-to-background ratio, as in panel (a) of Fig.~\ref{fig:plot_all_multiple}, models using the angle-averaging method cannot describe experiments with just a few detectors, giving rise to small mean p-values and large uncertainties on $\lambda$. For example, the 1-detector experiment has a very small mean p-value of 0.001, indicating an inconsistency between the data and the model. With increasing $n_D$, the experimental performance dramatically improves and quickly approaches the performance of the optimal case with $n_D\rightarrow \infty$. The confidence interval for a 50-detector experiment is only 7\% wider than the optimal case, and the mean p-value 0.47 is already very close to 0.5. The performance of a 150-detector experiment is almost identical to the optimal case, suggesting 150 detectors are more than sufficient to use Eq.~\ref{eq:average}. We conclude that experiments with more than about 15 identical detectors and a $\sigma_{E}=0.04\rm E$ energy resolution can use the averaging method safely with a modest increase of 10\% or so for the experimental uncertainty over the optimal case of $n_D\rightarrow \infty$. 

It is worth giving emphasis that even for the optimal case of $n_D\rightarrow \infty$, the experimental uncertainty and sensitivity on $\lambda$ is still about a factor of 3 to 4 worse than that in ES0, equivalent to a factor of 9 to 16 less exposure. This is due to the loss of individual detector information. Therefore, it is preferable to measure all absolute solar angles to ~$2\degree$ to~$4\degree$ as suggested in Sect.~\ref{sec:angle_precision}, whenever feasible. 

\section{Conclusions and Discussion}
If axions exist, they might be abundantly produced within the Sun and induce distinct photon signals via coherent Bragg-Primakoff conversions in solid state detectors. We performed simulation studies of a model of such signals in Ge crystals. We compared different experimental scenarios that correlate the possible axion signals in multiple Ge detectors, including a novel method of utilizing relative angle measurements between detectors. In this section we discuss conclusions derived from these studies. Although these conclusions are derived for the case of Ge detectors, they will apply to other solid state detectors. 

An angle-averaged axion signal pattern can be used to analyze an experiment, if there are more than about 15 identical detectors (Sec.~\ref{sec:averaging}). However, the sensitivity to axion-photon coupling $\lambda$ in this averaging method is a factor of 3 to 4 worse than measuring all detector angles precisely, if the latter is feasible. 

The main goal of the \mj\ \dem\ is to demonstrate an exceedingly low background in order to search for neutrinoless double beta decay (0$\nu\beta\beta$)~\cite{Abgrall2014, Xu2015}. This experiment will deploy more than thirty $^{76}$Ge-enriched HPGe detectors of roughly 0.9~kg each and about twenty natural detectors with almost identical mass of 0.6~kg each. The excellent energy resolution of these MJD detectors~\cite{Abgrall2016} allows a sensitivity that is comparable to what $\sigma_{E}=0.04\rm E$ can provide. According to our studies, the number of detectors in each of these two groups exceeds the minimum to apply the averaging analysis method described in Sec.~\ref{sec:averaging}. Furthermore, the axion sensitivity per kg of the heavier enriched detectors is similar to the natural detectors. Therefore, during the data analysis one could artificially reduce the effective mass of the heavier detectors to 0.6~kg by randomly eliminating roughly one third of all events. This would provide more than 50 identical detectors of 0.6~kg each. This analysis can be carried out without any significant experimental efforts beyond the 0$\nu\beta\beta$ program.

For some experiments, the relative angles between detectors can be measured well, but only poor knowledge of absolute angles to the Sun for each detector is available. The relative angle method in Sect.~\ref{sec:Relative_Angle} was demonstrated to be capable of improving experimental sensitivity to axions in comparison to the case of little angular information. This improvement is approximately 30\% for four detectors with a $\pm2\degree$ relative angle precision and it increases with the number of detectors deployed in an experiment.

The studies presented here were aimed at providing guidance for current and near-future experiments. A precision measurement of the absolute solar angle is best. With an energy resolution of $\sigma_E=0.04{\rm E}$, our study shows that a measurement of the absolute angle within 2$\degree-4\degree$ is sufficient. (Note, the CDMS experiment measured their crystal angles to approximately 3$\degree$~\cite{Ahmed2009}.) Importantly, with such an angular precision, an experiment can use the profile likelihood method with nominal critical values from a $\chi^2(1)$ distribution, avoiding extensive Monte Carlo studies. 

\section*{Acknowledgments}
This work was supported by the U.S. Department of Energy through the Los Alamos National Laboratory, Laboratory-Directed Research and Development Program. We thank Richard J. Creswick for useful discussions on the modeling of axion signals in HPGe detectors and Chenkun Wang for useful discussions on the statistical methods. We thank John F. Wilkerson, Jason Detwiler, and Christopher O'Shaughnessy for reading the manuscript and providing valuable comments. Computations supporting this project were in part performed on High Performance Computing systems at the University of South Dakota.

\appendix
\section{Statistical Discussions}\label{app:stat}
In this Appendix, we apply the profile likelihood method to Monte Carlo simulation data sets, demonstrate that the test statistic can deviate from standard $\chi^2$ distributions due to large systematic uncertainties and boundary constraints, and outline an adjustment to the critical values to ensure proper Frequentist coverages. Numerically obtaining proper critical values from Monte Carlo simulations is a common practice, such as in Ref.~\cite{Feldman1998}. There are alternative statistical procedures to analyze solar axion experiments with unknown detector angles, including the one prescribed by EDELWEISS~\cite{Armengaud2013}. 

\subsection{The Likelihood Function}
The likelihood function is modeled after the one derived in~\cite{Abdurashitov1999, Cleveland1983}. For an experiment with a total of $n_D$ detectors, the likelihood functions for individual detectors are multiplied to obtain the joint likelihood function, as,
\begin{equation}
\mathcal{L} =\prod_{j=1}^{n_D} \mathcal{L}^{(j)}_D~,
\label{eq:FullLikelihood}
\end{equation}
where $\mathcal{L}^{(j)}_D$ is for the $j^{th}$ detector. 

For each detector, in an energy region between $E_{lo}$ and $E_{hi}$ appropriate for axions, the data set can be described by the number of measured events in each detector. These events with time $t_i$ and energy $E_i$ arise either from background or an axion signal. The background is assumed to be constant in time and in energy, and therefore is described as a constant $b^{(j)}$ cts/(keV d). The signal rate depends on $t$ and $E$ and will be described as $\dot R(t,E)$, as given in Eq.~\ref{eq:compact}. Hence the expected total number of counts ($m$) in detector $j$ is:

\begin{equation}
m^{(j)}=\int_0^T\int_{E_{lo}}^{E_{hi}} [b^{(j)}+\dot R^{(j)}(t, E)]dtdE~,
\label{eq:totalcounts}
\end{equation}
where T is the stop time of the measurement. The likelihood for the $N^{(j)}$ observed events in the $j^{th}$ detector for this model is then:
\begin{equation}
\mathcal{L}_D^{(j)} = e^{-m^{(j)}}\prod_{i=1}^{N^{(j)}}[b^{(j)}+\dot R^{(j)}(t^{(j)}_i, E^{(j)}_i)]~.
\label{eq:likelihood}
\end{equation}
As stated in Sect.~\ref{sec:simu_overview}, the background in each detector was assumed to be proportional to detector mass ($M^{(j)}_D$),~\textit{i.e.} $b^{(j)}=\bar{b}M^{(j)}_D$, where $\bar{b}$ is an universal rate normalized in time, energy and mass. For experimental scenarios 0 and 1 described in Sec.~\ref{sec:Relative_Angle}, the absolute azimuthal angle for the $j^{th}$ detector ($\phi^{(j)}$) is implicit in its expected signal rate, $R^{(j)}$. For experimental scenario 2, the absolute angle for the $j^{th}$ detector can be expressed as $\phi^{(j)}=\Delta \phi^{(j)}+\phi^{(1)}$, where $\Delta \phi^{(j)}$ is the relative angle between the $j^{th}$ detector and the first detector. 

\subsection{Profile Likelihood}
Following~\cite{Rolke2005,Olive2014}, the profile likelihood $\mathcal{L}_{p}$ for a specific value of axion-photon coupling $\lambda_0$ is
\begin{equation}
\mathcal{L}_{p}(\lambda_0) = \frac{{\rm sup}\{\mathcal{L}(\lambda_0, \bar{b}, \vec\phi);\bar{b}, \vec\phi\}}{{\rm sup}\{\mathcal{L}(\lambda, \bar{b}, \vec\phi);\lambda, \bar{b}, \vec\phi\}}.
\label{eq:profilelikelihood}
\end{equation}
The joint likelihood $\mathcal{L}$ is maximized against all unknown parameters in the denominator, but only against the nuisance parameters of background and angles in the numerator. Maximizing $\mathcal{L}$ is equivalent to minimizing the negative log likelihood function, $-2~{\rm ln}\mathcal{L}$.

The signal rate expressed by Eq.~\ref{eq:compact} has a very complicated dependence on the detector azimuthal angle $\phi$ and the instantaneous position of the Sun. Therefore, the minimization of $-2~{\rm ln}\mathcal{L}$ against $\phi$ is performed on a $n_D$ dimension grid of $\phi$-angles. The effect of the step size of the angle grid was studied for Fig.~\ref{fig:angle}, where various step sizes ranging from $0.5\degree$ to $2.0\degree$ were tested and compared. Limited quantitative effect of the step size was observed but the conclusion was not affected. While a smaller step size would provide more precise quantitative results, its computational requirement would be overwhelming for multiple-detector studies. The data points in Fig.~\ref{fig:angle} used a $0.5\degree$ step size. With the exception of Fig.~\ref{fig:angle}, the default angle grid step size is $2.0\degree$, which is smaller than what was used by SOLAX (Fig.2 of~\cite{Avignone1998}) and COSME (Fig.4 of~\cite{Morales2002}) experiments when they scanned all possible detector angles. Given that the assumed systematic uncertainty on the relative angle is $\pm2.0\degree$ in Sect.~\ref{sec:Relative_Angle}, a smaller step size might be more preferable and remains a possible direction of improvement for future work.

The extent of the angle grid corresponds to the assumed systematic uncertainty on each angle. For a specific $\lambda_0$ and angle grid position, $-2~{\rm ln}\mathcal{L}$ is locally minimized against $\bar{b}$, using MINUIT~\cite{minuit} in ROOT~\cite{Brun97}. Maintaining the same $\lambda_0$, the local minima on the entire angle grid are then compared to find ${\rm sup}\{\mathcal{L}(\lambda_0, \bar{b}, \vec\phi);\bar{b}, \vec\phi\}$, which only depends on $\lambda_0$. Repeating this for a wide range of $\lambda_0$, the profile likelihood $\mathcal{L}_{p}(\lambda)$ is numerically obtained and the global minimum of $-2~{\rm ln}\mathcal{L}_{p}$ can be found at $\hat\lambda$, which is the best estimation of $\lambda$. 

The boundaries of a confidence interval (CI), $\hat\lambda_{up}$ and $\hat\lambda_{lo}$, can be found at values of $\lambda$, where $-2~{\rm ln}\mathcal{L}_p$ increases from the global minimum by a certain critical value. Let $D_{90}$ be the critical value associated with a 90\% confidence level (CL), and its nominal value would be 2.71 based on $\chi^2(1)$ distributed test statistic. However, to guarantee claimed coverage, we define a test statistic as below and $D_{90}$ is then calculated based on observed distributions of the test statistics in the simulation ensembles. $D_{90}$ often differs from the nominal value of 2.71. For clarification, $D_{90}$ obtained from Monte Carlo simulation is called the adjusted critical value.

\subsection{Test Statistic and Adjusted Critical Value}
A test statistic $D$ is defined as:
\begin{eqnarray}
\label{eq:Dtest}
D&\equiv&-2{\rm ln}\mathcal{L}_{p}(\lambda_{true}) \\ \nonumber
&=&-2{\rm ln}\frac{{\rm sup}\{\mathcal{L}(\lambda_{true}, \bar{b}, \vec\phi);\bar{b}, \vec\phi\}}{{\rm sup}\{\mathcal{L}(\lambda, \bar{b}, \vec\phi);\lambda, \bar{b}, \vec\phi\}}
\end{eqnarray}
where $\lambda_{true}$ is the value used to generate simulation data sets. Let the distribution of $D$ in an ensemble of simulations be $P_D$. Adjusted critical value $D_{90}$ can be numerically obtained by requiring
\begin{equation}
\frac{\int_0^{D_{90}}P_DdD}{\int_0^{+\infty} P_DdD}=0.90~.
\label{eq:likelihood}
\end{equation}
$D_{90}$ replaces the nominal value of 2.71 in the calculation of a 90\% CI. If $\lambda=0$ is reached but $-2{\rm ln}\mathcal{L}_{p}$ increases less than $D_{90}$, then $\hat\lambda_{lo}=0$ and $\hat\lambda_{up}$ is unaffected. The main concern here is how much ${P_D}$ deviates from $\chi^2$ distributions (see below).

\subsection{Behavior of Test Statistic $D$}
\begin{figure}
\includegraphics[width=0.5\textwidth]{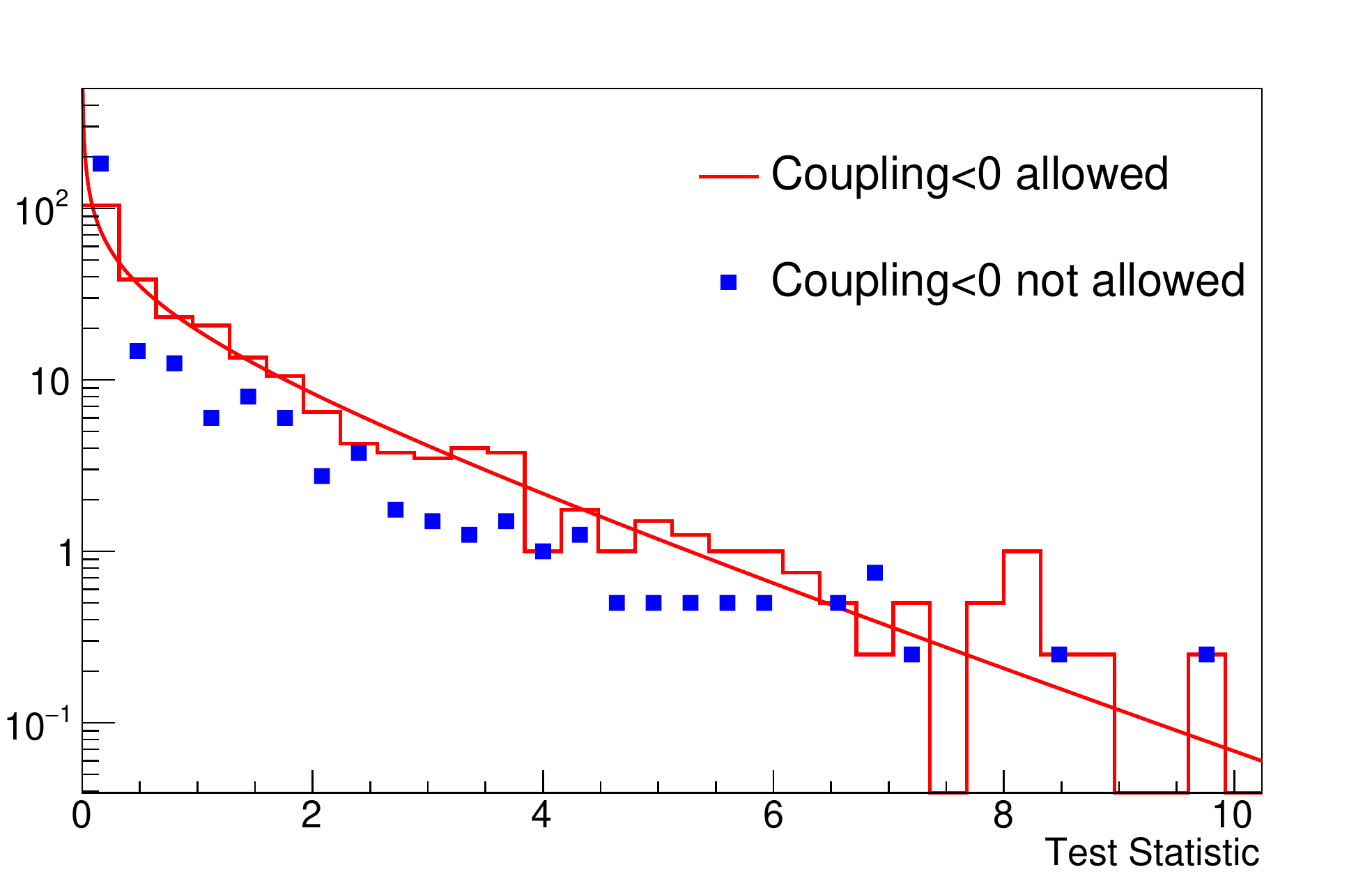}
\caption{Test statistic distribution for $\lambda_{true}=0,~\bar{b}=0.1$~cts/(keV kg d) and $\phi_{mea}=\phi_{true}$ for a 4-detector experiment. The red histogram is for the case of no restrict on $\lambda$, which can be described by $\chi^2(1)$ (red curve). Blue filled squares show the case where $\lambda$ is restricted to be non-negative, and the distribution is slightly narrower than $\chi^2(1)$.}
\label{fig:2a} 
\end{figure}

When no physical boundary exists and the nuisance parameters are either known or can be accurately profiled out from data, the probability of axion-induced signals as a function of the single parameter $\lambda$ is fixed by Eq.~\ref{eq:compact}. In this case, according to Wilks' theorem~\cite{Wilks1938}, the test statistic $D$ asymptotically follows a $\chi^2$ distribution with 1 degree of freedom (DoF), \textit {i.e.} $P_D \propto\chi^2(1)$. Changing the value of $-2{\rm ln}\mathcal{L}_{p}$ by 2.71 approximately produces a 90\% CI, given that the $\chi^2(1)$ cumulative probability at 2.71 is 90\%. The same approximation also results in the simply correspondence between a $1\sigma$ uncertainty and $\Delta{\rm ln\mathcal{L}}=1/2$~\cite{Rolke2005}, as well as powerful analytic formulas that can be used to replace extensive simulation studies~\cite{Cowan2011}. This approximation is observed in certain simulations, as the red histogram in Fig.~\ref{fig:2a} shows. 

However, physical boundaries and/or insufficient knowledge on parameters may affect test statistic distribution. Wilks' theorem requires parameters are not on the boundaries, and the impact of physical boundaries has been widely recognized in the literature, for example Ref.~\cite{Feldman1998}. This is demonstrated by the blue squares in Fig.~\ref{fig:2a}, where the axion-photon coupling is restricted to be non-negative. The observed test statistic distribution is narrower than $\chi^2(1)$ and over-coverage would occur if the nominal critical value 2.71 is used for a 90\% CI. 

\begin{figure}
\includegraphics[width=0.5\textwidth]{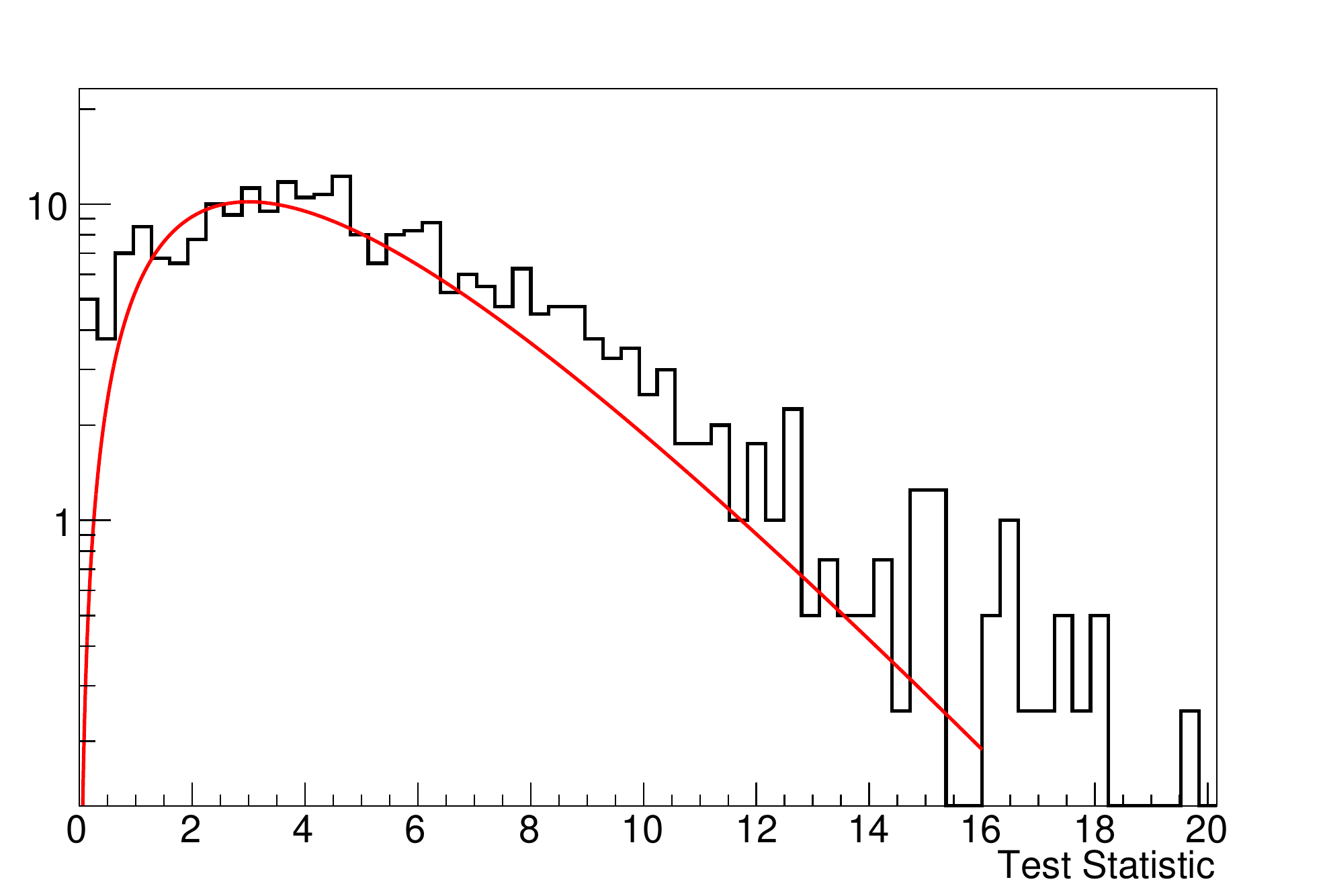}
\caption{Test statistic distribution (black histogram) when there are no axion signals with $\lambda_{true}=0,~\bar{b}=0.1$~cts/(keV kg d) and $\phi_{mea}\in(\phi_{true}-7.5{\degree},  \phi_{true}+7.5{\degree})$ for a 4-detector experiment, analyzed using Experimental Scenario 1. The red curve is a $\chi^2(5)$ distribution for comparison.}
\label{fig:2c} 
\end{figure}

Insufficient knowledge is a broad condition, and specific example are discussed below using the Monte Carlo simulations with $n_{D}=4$ detectors from Sect.~\ref{sec:Relative_Angle}. When $\lambda_{true}=0$, there is no axion signal to help profile out detector angles and the background has no angle dependence. The distribution of test statistic D solely depends on the allowed angle range in the analysis, which is 15{\degree} in Experimental Scenario 1. As shown in Fig.~\ref{fig:2c}, the observed test statistic approximately follows $\chi^2(5)$, and the adjusted critical value $D_{90}$ is numerically determined to 9.64. The $\chi^2(5)$ cumulative probability at 9.64 is 91$\%$, supporting $\chi^2(5)$ being a good analytic approximation. This can be understood as 15$\degree$ angular uncertainty is large enough that the $n_{D}$ detector angles are effectively unknown and unconstrained. Due to $\lambda_{true}=0$, no information on the $n_{D}$ angles can be gained from the data either. In this extreme case of no knowledge on $n_{D}$ nuisance parameters, the test statistic approximates a $\chi^2$ distribution with $1+n_{D}$ DoF. 

However, the ability to profile out detector angles depends on the strength of the hypothetical axion signals, possibly lessening the insufficient knowledge condition. A further complication is that the detector angles are more tightly constrained in Experimental Scenario 2 with good relative angle measurements within 4{\degree}. It is therefore difficult to predict the general behavior of the test statistic analytically. Instead, the $D$ distribution and the corresponding critical values can be mapped out from extensive simulations, as done here. 

\begin{figure}
\includegraphics[width=0.50\textwidth]{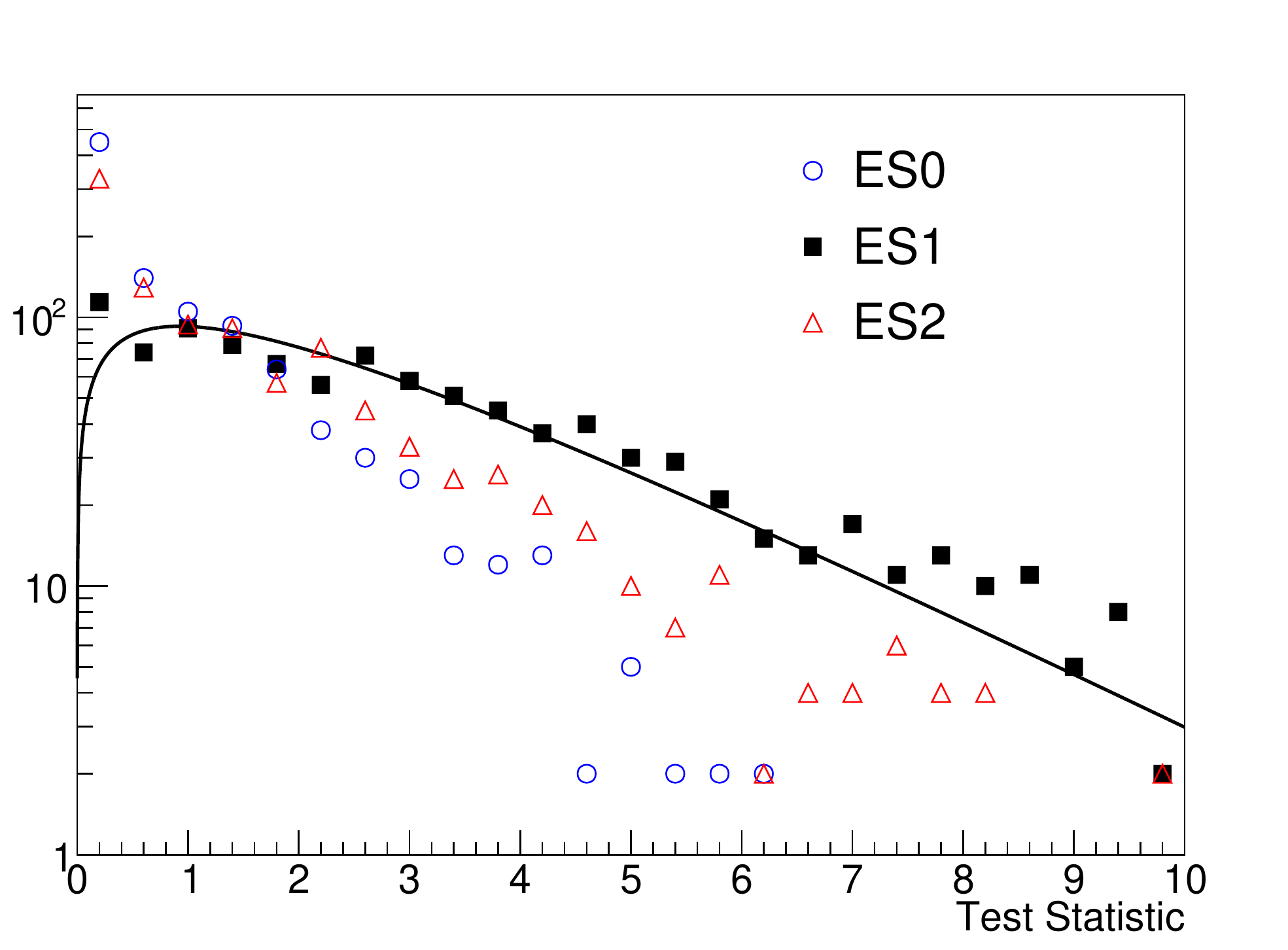}
\caption{Test statistic distribution with $\lambda_{true}=1\times10^{-4},\bar{b}=0.1$~cts/(keV kg d) for a 4-detector experiment. The black curve is a $\chi^2(2.9)$ distribution for comparison. Experimental scenario 1 (black filled squares) with $\phi_{mea}\in(\phi_{true}-7.5{\degree},  \phi_{true}+7.5{\degree})$ has the most broad distribution, but still narrower than $\chi^2(5)$, in contrast to Fig.~\ref{fig:2c}.}
\label{fig:3} 
\end{figure}

For example, Fig.~\ref{fig:3} compares the test statistic distribution in the case of $\lambda_{true}=1\times10^{-4}, \bar{b}=0.1$ cts/(keV kg d) for three different experimental scenarios. The $D$ distribution for ES1 is narrower than a $\chi^2(5)$ distribution, thanks to the small but non-zero axion signals in the data introducing a limited sensitivity on angles. Replacing insufficient knowledge on absolute angles with good knowledge on relative angles, ES2 has a much narrower distribution than ES1, but still wider than ES0 with exact angular information. In simulations with $\lambda_{true}=0.0005$, axion signals in the data are large enough to profile out detector angles accurately in all three experimental scenarios, and all test statistic distributions approximate $\chi^2(1)$ regardless of initial angular knowledge.

There exists a large volume of discussions in the literature on treating confidence intervals with systematic uncertainties. See for example Refs.~\cite{Cousins1992, Conard2003, Heinrich2007, Cousins2008}. In the solar axion experiments considered here, both the auxiliary measurements and the hypothetical axion signals contribute to the knowledge on detector angles in complicated ways. In our studies, we represented auxiliary measurement constraints in the generation and analysis of Monte Carlo simulations, and let the likelihood profiling process handle the constraints from data, in an attempt to properly map out the restricted unknown parameter space and numerically obtain proper critical values. It was verified this procedure guarantees the claimed coverage in Monte Carlo simulations from Sect.~\ref{sec:Relative_Angle}, for both zero and non-zero $\lambda_{true}$.


\bibliographystyle{elsarticle-num}
\bibliography{SolarAxion_Xu_Elliott}






\end{document}